\documentclass[a4paper,10pt]{elsarticle}
\usepackage{framed,multirow}
\usepackage[utf8]{inputenc}
\usepackage{bm}
\usepackage{amsmath}
\usepackage{mathrsfs}
\usepackage{graphicx}
\usepackage{epsfig, setspace}
\usepackage{url}
\usepackage{graphicx, transparent, color}
\usepackage{algorithm} 
\usepackage{algorithmicx}
\usepackage{etex}
\usepackage{siunitx}
\usepackage[bookmarks=false]{hyperref}
\usepackage{amsmath}
\usepackage{xcolor}
\usepackage{rotating} 
\usepackage{pdflscape} 
\usepackage{silence}
\usepackage{ulem,cancel}
\usepackage[utf8]{inputenc}

\usepackage{graphicx}
\usepackage{float}
\usepackage{subfigure}
\usepackage{subfigmat}

\usepackage{tikz}
\usepackage{capt-of}
\usepackage{setspace,lipsum}
\usepackage{lineno}
\usepackage{dirtytalk}

\usepackage[top=18truemm,bottom=18truemm,left=20truemm,right=20truemm]{geometry}
\DeclareFontFamily{U}{mathx}{}
\DeclareFontShape{U}{mathx}{m}{n}{<-> mathx10}{}
\DeclareSymbolFont{mathx}{U}{mathx}{m}{n}
\DeclareMathAccent{\widehat}{0}{mathx}{"70}
\DeclareMathAccent{\widecheck}{0}{mathx}{"71}

\begin{document}
\begin{frontmatter}
\title{On the role of spectral properties of viscous flux discretization for flow simulations on marginally resolved grids}
\author[]{Amareshwara Sainadh Chamarthi \cortext[cor1]{Corresponding author. \\ 
E-mail address: skywayman9@gmail.com (Amareshwara Sainadh  Ch.).}}
\author[]{Hemanth Chandravamsi K}
\author[]{Natan Hoffmann}
\author[]{Sean Bokor} 
\author[]{Steven H.\ Frankel}
\address{Faculty of Mechanical Engineering, Technion - Israel Institute of Technology, Haifa, Israel}

\begin{abstract}

In this note, the importance of spectral properties of viscous flux discretization in solving compressible Navier-Stokes equations for turbulent flow simulations is discussed. We studied six different methods, divided into two different classes, with poor and better representation of spectral properties at high wavenumbers. Both theoretical and numerical results have revealed that the method with better properties at high wavenumbers, denoted as $\alpha$-damping type discretization, produced superior solutions compared to the other class of methods. The proposed compact $\alpha$-damping method converged towards the direct numerical simulation (DNS) solution at lower grid resolution compared with the other class of methods and is, therefore, a better candidate for high fidelity large-eddy simulations (LES) and DNS studies.

\end{abstract}


\end{frontmatter}

\section{Introduction}\label{sec1}

\textit{\say{It has been long assumed that the accuracy of the convective terms is critical for resolving the turbulence and the accuracy of viscous terms is much less important}} \cite{debonis2013solutions}. Thus far, the conventional wisdom has been that viscous fluxes do not play a critical role in resolving turbulence. However, in this note, we demonstrate through theoretical and numerical analysis that this is not the case. Therefore, it is important to isolate the effects of discretization for the convective and viscous terms of the Navier-Stokes equations. With the advent of the kinetic energy and entropy preserving approach \textcolor{black}{\cite{chandrashekar2013kinetic, kuya2018kinetic, pirozzoli2010generalized,kennedy2008reduced}}, it is possible to isolate the effect of the viscous flux discretization on flow simulations. We compared several viscous flux discretizations to demonstrate the importance of their spectral properties on the solution quality of flow simulations. To the best of the authors' knowledge, such an analysis in the literature has not been formally addressed thus far. \\

For compressible flow simulations, the inviscid fluxes in the compressible Navier-Stokes simulations are typically discretized using upwind schemes \cite{Shu1997,Hu2010}, which have inherent numerical dissipation that is necessary for flows with shocks and material discontinuities. The inherent numerical dissipation of upwind schemes may sometimes over-approximate the physical dissipation of the flow, which makes it difficult to assess the contribution and effect of the numerical viscous flux discretization. Moreover, for certain cases not involving discontinuities, upwind schemes are still employed since central schemes can lead to numerical instabilities. While upwind schemes may help avoid blowup in these cases, excessive dissipation may damp turbulent flow features. Filtering is another method to contain the instabilities \cite{visbal2002use}, but this too may cause excessive dissipation. \\

Besides upwind schemes and filtering, other methods exist to stabilize numerical simulations while minimizing dissipation. For instance, Nagarajan et al. \cite{nagarajan2003robust} used the split form approach of Blaisdell \cite{blaisdell1991numerical} for the inviscid fluxes. For the viscous fluxes, they used the non-conservative form by expanding them in Laplacian form and proposed a robust viscous flux discretization using compact finite-difference schemes \cite{lele1992compact}. They suggested that a scheme with good spectral properties at high wavenumbers is essential for stable simulations. However, regarding the inviscid flux discretization, the split form approach of Blaisdell \cite{blaisdell1991numerical} is not kinetic energy preserving, making it difficult to isolate the effects of the viscous flux discretization. While the approaches mentioned above are frequently utilized, one method that uses central schemes that is stable and non-dissipative is the kinetic energy and entropy preserving (KEEP) scheme proposed by Kuya and Kawai \cite{kuya2018kinetic}. Lamballais et al. \cite{lamballais2011straightforward,dairay2017numerical,lamballais2021viscous} used the kinetic energy preserving approach for the inviscid fluxes ensuring that kinetic energy dissipation only occurred due to the viscous fluxes in their work. However, they solve the incompressible Navier-Stokes equations, whereas, in this paper, the compressible Navier-Stokes equations are solved. Also, the motivation of Lamballais and his collaborators was to propose a discretization for viscous fluxes such that it acts like a filter that provides extra dissipation instead of a subgrid-scale model. In contrast, we seek to demonstrate how proper viscous flux discretization is required to obtain physically consistent results on under-resolved grids. \\

Discretization of viscous fluxes for the compressible Navier-Stokes equations, specifically in the context of the KEEP scheme, has not been thoroughly studied. A fourth- and sixth-order discretization of the viscous fluxes in conservative form was proposed by Shen et al. \cite{shen2009high,shen2010large}. Recently, Chamarthi et al. \cite{chamarthi2022} have shown that the discretization approach of Shen et al. \cite{shen2010large} will lead to odd-even decoupling for compressible flow simulations involving shock waves. Chamarthi et al. \cite{chamarthi2022} proposed a conservative sixth-order viscous scheme by defining a numerical flux with low-order consistent and damping terms involving free parameters that determine the order of accuracy and spectral properties of the discretization. This approach, known as the $\alpha$-damping approach, was first proposed by Nishikawa \cite{nishikawa:AIAA2010}. In his seminal paper, Nishikawa proposed viscous flux discretization approaches for various numerical methods, including the Discontinuous Galerkin, Finite Volume, and Spectral Volume methods. 

\textcolor{black}{In reference \cite{chamarthi2022}, apart from the derivation of a sixth-order  $\alpha$-damping viscous scheme, the spectral properties of the sixth-order scheme proposed by Shen et al. \cite{shen2010large} are also analyzed (which was not carried before to the best of the authors' knowledge). It was shown in reference \cite{chamarthi2022} that the Shens' scheme lacks high-frequency damping and will lead to odd-even decoupling. Furthermore, in Ref. \cite{chamarthi2022gradient}, Chamarthi has extended the  $\alpha$-damping approach to compact finite difference schemes \cite{lele1992compact} and derived fourth-order viscous schemes with superior spectral properties than the sixth-order Shens' scheme. Also, the objective of Ref. \cite{chamarthi2022gradient} was to propose a new approach such that the gradients, once computed, can be used for both inviscid and viscous fluxes. The test cases considered in both papers \cite{chamarthi2022} and \cite{chamarthi2022gradient} involve discontinuities and require an upwind scheme. Upwind schemes will add numerical dissipation, and it is difficult to understand the effect of a viscous scheme. In order to understand the viscous flux discretization effect, this paper uses kinetic energy and entropy preserving approach to model the inviscid fluxes. This paper proposes a sixth-order $\alpha$-damping method based on a compact finite-difference scheme with excellent spectral properties (better than the schemes presented in Ref. \cite{chamarthi2022gradient}). We come to show that the spectral properties of the viscous flux discretization have a significant effect on solution quality using under-resolved grids.}


This note is organized as follows. Section \ref{sec2} introduces the governing equations and the discretization of inviscid fluxes. Section \ref{sec3} presents various schemes' viscous flux discretization and their corresponding spectral properties. Numerical results are presented in Section \ref{sec4}. Finally, in Section \ref{sec5}, we provide conclusions.



\section{Computational Approach}\label{sec2}

\subsection{Governing Equations}

In this study, the three-dimensional compressible Navier-Stokes equations in conservative form are solved in Cartesian coordinates:

\begin{equation} \label{eqn:cns}
    \frac{\partial \mathbf{U}}{\partial t} + \frac{\partial \mathbf{F}^c}{\partial x} + \frac{\partial \mathbf{G}^c}{\partial y} + \frac{\partial \mathbf{H}^c}{\partial z} + \frac{\partial \mathbf{F}^v}{\partial x} + \frac{\partial \mathbf{G}^v}{\partial y} + \frac{\partial \mathbf{H}^v}{\partial z} = 0,
\end{equation}

\noindent where $t$ is time and $(x,y,z)$ are the Cartesian coordinates. $\mathbf{U}$ is the conserved variable vector, and $\mathbf{F}^c$, $\mathbf{G}^c$, and $\mathbf{H}^c$ are the inviscid flux vectors defined as:

\begin{subequations}\label{eqn:invFluxes}
    \begin{align}
        \mathbf{U} &= \left[ \rho, \rho u, \rho v, \rho w, \rho E \right]^\text{T}, \\
        \mathbf{F}^c &= \left[ \rho u, \rho u^2 + p, \rho u v, \rho u w, \rho u H \right]^\text{T}, \\
        \mathbf{G}^c &= \left[ \rho v, \rho v u, \rho v^2 + p, \rho v w, \rho v H \right]^\text{T}, \\
        \mathbf{H}^c &= \left[ \rho w, \rho w u, \rho w v, \rho w^2+p, \rho w H \right]^\text{T},
    \end{align}
\end{subequations}

\noindent where $\text{T}$ denotes transpose, $\rho$ is the density, $u$, $v$, and $w$ are the velocities in the $x$, $y$, and $z$ directions, respectively, $p$ is the pressure, $E = e + \left(u^2 + v^2 + w^2 \right)/2$ is the specific total energy, and $H = E + p/\rho$ is the specific total enthalpy. The equation of state is for a calorically perfect gas so that $e = p/ \left[ \rho (\gamma-1) \right]^{-1}$ is the internal energy, where $\gamma = \mathrm{c_p}/\mathrm{c_v}$ is the ratio of specific heats with $\mathrm{c_p}$ as the isobaric specific heat and $\mathrm{c_v}$ as the isochoric specific heat. $\mathbf{F}^v$, $\mathbf{G}^v$, and $\mathbf{H}^v$ are the viscous flux vectors defined as:

\begin{subequations}\label{eqn:viscFluxes}
    \begin{align}
        \mathbf{F}^v &= -\left[0, \tau_{x x}, \tau_{x y}, \tau_{x z}, u \tau_{x x}+v \tau_{x y}+w \tau_{x z}-q_{x}\right]^{\text{T}}, \\
        \mathbf{G}^v &= -\left[0, \tau_{y x}, \tau_{y y}, \tau_{y z}, u \tau_{y x}+v \tau_{y y}+w \tau_{y z}-q_{y}\right]^{\text{T}}, \\
        \mathbf{H}^v &= -\left[0, \tau_{z x}, \tau_{z y}, \tau_{z z}, u \tau_{z x}+v \tau_{z y}+w \tau_{z z}-q_{z}\right]^{\text{T}},
    \end{align}
\end{subequations}

\noindent where the normal stresses are defined as:

\begin{subequations}\label{visc}
    \begin{gather}
        \tau_{xx} = 2 \hat{\mu} \frac{\partial u}{\partial x} + \hat{\lambda} \left(\frac{\partial u}{\partial x} + \frac{\partial v}{\partial y} + \frac{\partial w}{\partial z} \right),
        \tag{\theequation a-\theequation b}
        \quad
        \tau_{yy} = 2 \hat{\mu} \frac{\partial v}{\partial y} + \hat{\lambda} \left(\frac{\partial u}{\partial x} + \frac{\partial v}{\partial y} + \frac{\partial w}{\partial z} \right), \\
        \tau_{zz} = 2 \hat{\mu} \frac{\partial w}{\partial z} + \hat{\lambda} \left(\frac{\partial u}{\partial x} + \frac{\partial v}{\partial y} + \frac{\partial w}{\partial z} \right),
        \tag{\theequation c}
    \end{gather}
\end{subequations}

\noindent where $\hat{\mu} = \mu/\mathrm{Re}$ is the scaled dynamic viscosity as a result of non-dimensionalization and Stokes' hypothesis is assumed so that $\hat{\lambda} = -\frac{2}{3} \hat{\mu}$. $\mathrm{Re} = \rho_{\infty} u_{\infty} L_{ref}/\mu_{\infty}$ is the Reynolds number where the $\infty$ subscript denotes a freestream value. The shear stresses are defined as:

\begin{subequations}
    \begin{gather}
        \tau_{xy} = \tau_{yx} = \hat{\mu} \left(\frac{\partial u}{\partial y} + \frac{\partial v}{\partial x} \right),
        \quad
        \tau_{yz} = \tau_{zy} = \hat{\mu} \left(\frac{\partial v}{\partial z} + \frac{\partial w}{\partial y} \right),
        \quad
        \tau_{xz} = \tau_{zx} = \hat{\mu} \left(\frac{\partial u}{\partial z} + \frac{\partial w}{\partial x} \right),
        \tag{\theequation a-\theequation c}
    \end{gather}
\end{subequations}

\noindent and the heat fluxes are:

\begin{subequations}
    \begin{gather}
        q_{x} = -\hat{\kappa} \frac{\partial T}{\partial x},
        \quad
        q_{y} = -\hat{\kappa} \frac{\partial T}{\partial y},
        \quad
        q_{z} = -\hat{\kappa} \frac{\partial T}{\partial z},
        \tag{\theequation a-\theequation c}
    \end{gather}
\end{subequations}

\noindent where $\hat{\kappa} = \mu \left( \mathrm{Ma}^2 \mathrm{Re}(\gamma-1)\mathrm{Pr} \right)^{-1}$ is the scaled thermal conductivity, where $\mathrm{Ma} = u_{\infty} \left( \gamma R_{gas} T \right)^{-1/2}$ is the Mach number, $\mathrm{Pr}$ is the Prandtl number, $T$ is the temperature, and $R_{gas}$ is the universal gas constant. The equations are non-dimensionalized using the freestream density $\rho_{\infty}$, the freestream velocity $u_{\infty}$, reference length $L_{ref}$, the freestream temperature $T_{\infty}$, and the freestream dynamic viscosity $\mu_{\infty}$ such that the temperature is related to pressure and density via $ p = \rho T \left( \gamma \mathrm{Ma}^2 \right)^{-1}$. \\

\subsection{Inviscid Flux Discretization}\label{sec2.1}

In this paper, we seek to isolate the effect of the viscous flux discretization. As such, we use an inviscid flux discretization that is non-dissipative and stable. One such method is that of Kuya and Kawai \cite{kuya2021high}, in which they recast the inviscid terms of the Navier-Stokes equations in KEEP form. Since this method is kinetic energy preserving, it is non-dissipative, and as an added important factor, it is more numerically stable due to the discrete preservation of entropy. As a result, we can monitor the quality of the viscous flux discretization. In Cartesian coordinates, the recast equations can be expressed as:

\begin{subequations}
    \begin{align}
        &\frac{\partial \rho}{\partial t} + \frac{1}{2} \left( \frac{\partial \rho u_{j}}{\partial x_{j}} + \rho \frac{\partial u_{j}}{\partial x_{j}} + u_{j} \frac{\partial \rho}{\partial x_{j}} \right) = 0, 
        \tag{\theequation a} 
        \\
        &\frac{\partial \rho u_{i}}{\partial t} + \frac{1}{4}\left(\frac{\partial \rho u_{i} u_{j}}{\partial x_{j}} + u_{i} \frac{\partial \rho u_{j}}{\partial x_{j}} + u_{j} \frac{\partial \rho u_{i}}{\partial x_{j}} + \rho \frac{\partial u_{i} u_{j}}{\partial x_{j}} + \rho u_{j} \frac{\partial u_{i}}{\partial x_{j}} + \rho u_{i} \frac{\partial u_{j}}{\partial x_{j}} + u_{i} u_{j} \frac{\partial \rho}{\partial x_{j}}\right) + \frac{\partial p}{\partial x_{i}} = 0,
        \tag{\theequation b} 
        \\
        \begin{split}
            &\frac{\partial E}{\partial t} + \frac{1}{2} \left( u_{i} \frac{\partial \rho u_{i} u_{j} / 2}{\partial x_{j}} + \frac{\rho u_{i} u_{j}}{2} \frac{\partial u_{i}}{\partial x_{j}} + \rho u_{i} \frac{\partial u_{i} u_{j} / 2}{\partial x_{j}} + \frac{u_{i} u_{j}}{2} \frac{\partial \rho u_{i}}{\partial x_{j}} \right) + \\
            &\frac{1}{4} \left( \frac{\partial \rho e u_{j}}{\partial x_{j}} + u_{j} \frac{\partial \rho e}{\partial x_{j}} + e \frac{\partial \rho u_{j}}{\partial x_{j}} + \rho \frac{\partial e u_{j}}{\partial x_{j}} + \rho e \frac{\partial u_{j}}{\partial x_{j}}+\rho u_{j} \frac{\partial e}{\partial x_{j}}+ e u_{j} \frac{\partial \rho}{\partial x_{j}}\right) + \left(u_{j} \frac{\partial p}{\partial x_{j}} + p \frac{\partial u_{j}}{\partial x_{j}} \right) = 0,
        \end{split}
        \tag{\theequation c}
    \end{align}
\end{subequations}

\noindent where the indices are used according to Einstein notation here. Considering only the $x$-direction, these equations are spatially discretized on a uniform grid with $N$ cells on a type 2 (\textcolor{black}{cell-centered grid}) \cite{laney1998computational} spatial grid $x \in [x_a,x_b]$. The grid spacing is:

\[
    \Delta x = (x_b-x_a)/N
\]

\noindent The cell centers are located at:

\[
    x_j = x_a + (j - 1/2) \Delta x \quad \text{for } j = 1, 2, \dotsc, N
\]

\noindent The cell-interfaces are located at:

\[
    x_{j+\frac{1}{2}} \quad \text{for } j = 0, 1, \dotsc, N
\]

\noindent All gradients needed for the KEEP form of the conservation equations are computed with a sixth-order explicit central difference scheme. For example, the first derivative of streamwise velocity is computed as:

\begin{equation}
    \left( \frac{\partial u}{\partial x} \right)_j = \frac{1}{\Delta x} \left( -\frac{1}{60} u_{j-3} + \frac{3}{20} u_{j-2} - \frac{3}{4} u_{j-1} + \frac{3}{4} u_{j+1} - \frac{3}{20} u_{j+2} + \frac{1}{60} u_{j+3} \right),
\end{equation}

\noindent where the subscripts refer to cell-centers and $\Delta x$ is the grid spacing in the $x$-direction. This concludes the presentation of the KEEP scheme. \\

The implementation of the KEEP approach employed in the flow solver is verified by solving the three-dimensional inviscid Taylor-Green vortex (TGV) problem. The initial conditions for this test case are: 

\begin{equation}\label{itgv}
    \begin{pmatrix}
        \rho \\
        u \\
        v \\
        w \\
        p
    \end{pmatrix}
    =
    \begin{pmatrix}
        1 \\
        \sin{x} \cos{y} \cos{z} \\
        -\cos{x} \sin{y} \cos{z} \\
        0 \\
        100 + \frac{\left( \cos{(2z)} + 2 \right) \left( \cos{(2x)} + \cos{(2y)} \right) - 2}{16}
    \end{pmatrix}.
\end{equation}

\begin{figure}[h!]
\centering
\subfigure[\textcolor{black}{Normalized Kinetic energy}.]{\includegraphics[width=0.48\textwidth]{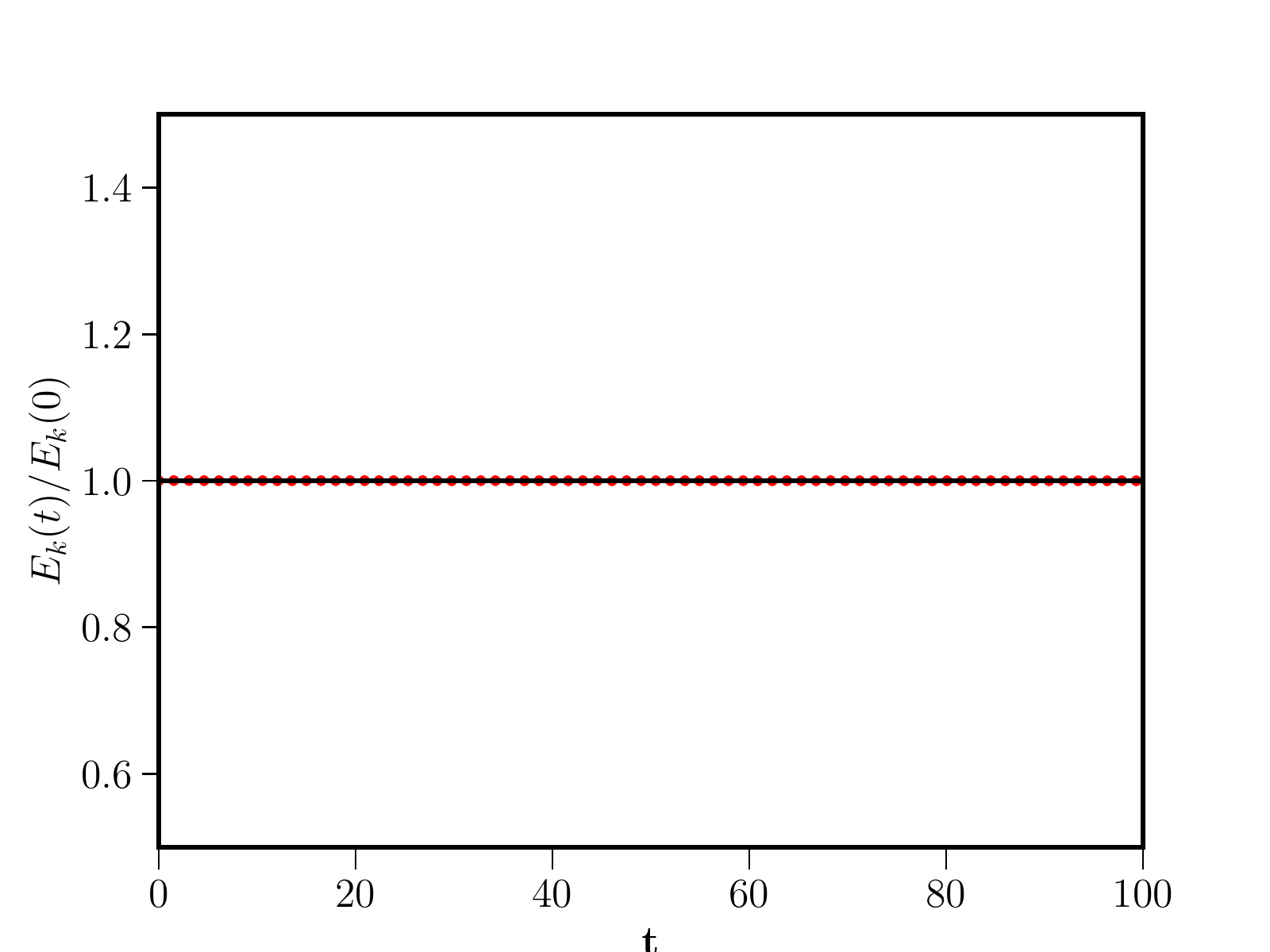}
\label{fig:TGV_KE}}
\subfigure[\textcolor{black}{Normalized Enstrophy.}]{\includegraphics[width=0.48\textwidth]{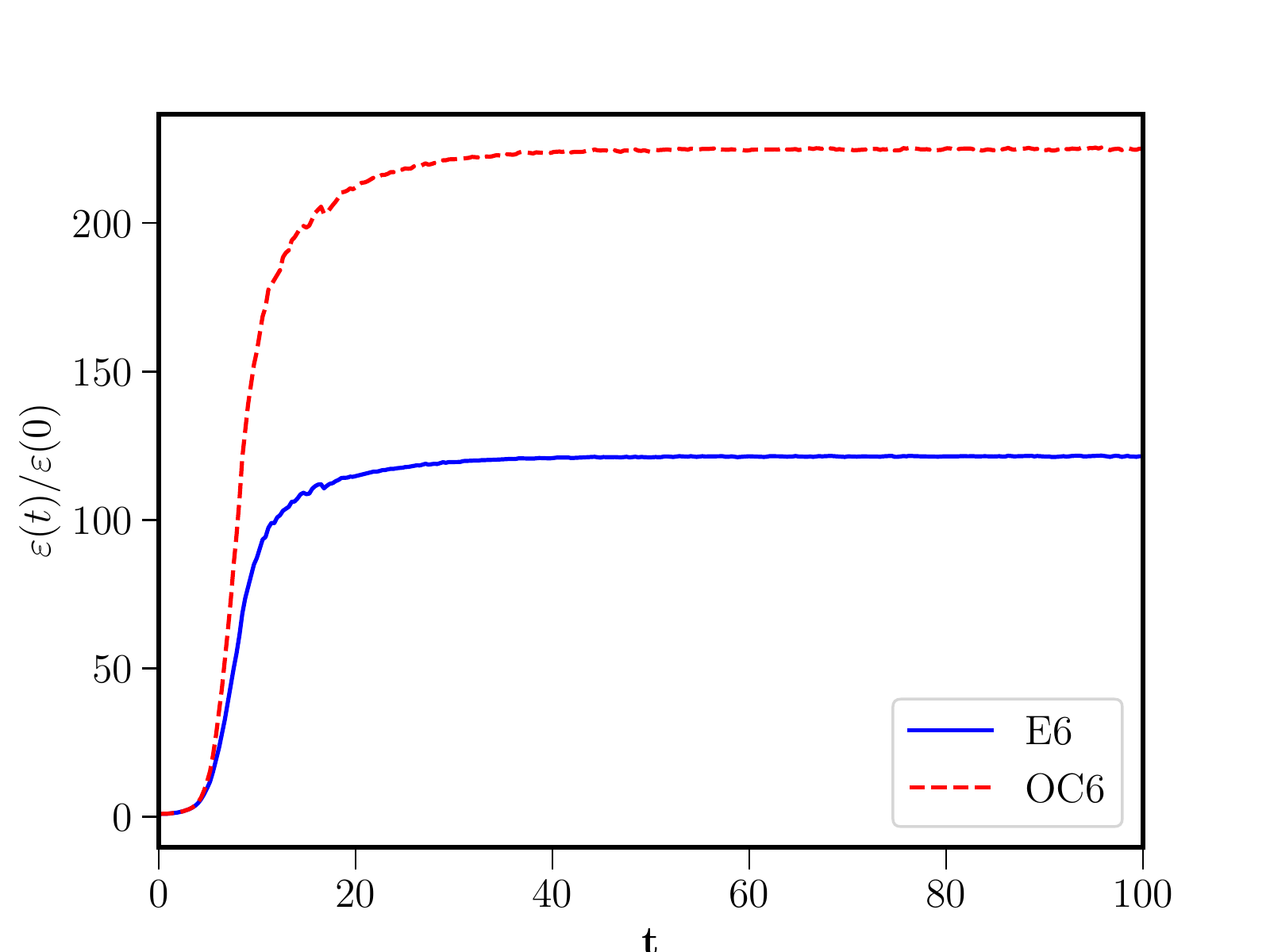}
\label{fig:TGV_ens}}
\caption{Normalized kinetic energy and enstrophy for inviscid TGV. Solid black line: exact solution. Red circles: numerical kinetic energy. Red dashed lines: enstrophy with OC6. Blue solid line: enstrophy with E6.}
\label{fig_TGV}
\end{figure}

\noindent The simulations are carried out on a domain of size $x,y,z \in [0,2\pi]$ with periodic boundary conditions applied for all boundaries. The simulations are performed until time $t=100$ on a grid size of $64^3$ with the specific heat ratio as $\gamma=5/3$. The mean pressure is relatively large, so the problem can be considered incompressible. \\

The volume-averaged kinetic energy and enstrophy evolution are shown in Fig. \ref{fig_TGV} up to $t=100$. These quantities were respectively computed as:

\begin{equation}
    E_{k}=\frac{1}{ L_x L_y L_z } \int_{0}^{Lz} \int_{0}^{Ly} \int_{0}^{Lx} \rho \frac{u^2 + v^2 +w^2}{2} \mathrm{~d}x\mathrm{~d}y\mathrm{~d}z,
\end{equation}

\begin{equation}
    \mathcal{E}=\frac{1}{ L_x L_y L_z } \int_{0}^{Lz} \int_{0}^{Ly} \int_{0}^{Lx} \rho \frac{ \mathbf{\Omega} \cdot \mathbf{\Omega}}{2} \mathrm{~d}x\mathrm{~d}y\mathrm{~d}z,
\end{equation}

\noindent where $\mathbf{\Omega} = \left( \Omega_x, \Omega_y, \Omega_x \right)$ represents the vorticity vector and it's components. The velocity derivatives required to compute the enstrophy are computed by the sixth-order compact finite difference scheme, denoted as OC6 (Equation (\ref{eqn-OC6})), which is presented in the next section. The use of high-order derivatives for the computation of enstrophy is important since vorticity is directly dependent on velocity gradients. Readers can refer to Appendix F of the article by Subramaniam et al. \cite{subramaniam2019high} regarding the effect of the numerical scheme used in post-processing for the computation of velocity gradients to compute enstrophy. All else equal, if the computation of enstrophy is done with two different difference schemes, there can be vastly different results. Indeed, observing Fig. \ref{fig:TGV_ens}, the enstrophy computed using an explicit sixth-order central difference scheme (E6) is far less than that computed using OC6. And as a testament to the kinetic energy preservation, Fig. \ref{fig:TGV_KE} displays no kinetic energy dissipation from the inviscid flux discretization. 

\section{Viscous Flux Discretization}\label{sec3}

As explained in the introduction, we present two different classes of viscous flux discretization: $\alpha$-damping methods and non-$\alpha$-damping (NAD) methods. The discretization and spectral properties of these methods are presented below.

\subsection{$\alpha$-Damping Approach}

First, we present the $\alpha$-damping approach, initially proposed by Nishikawa in Ref. \cite{nishikawa:AIAA2010}. In this approach, the viscous fluxes are discretized in conservative form. In this paper, we derive a new $\alpha$-damping method using a compact finite-difference scheme \cite{lele1992compact,kim1996optimized} that has superior spectral properties compared to the sixth-order explicit scheme proposed in \cite{chamarthi2022}. The details are presented here. For simplicity and without loss of generality, we will consider a one-dimensional scenario where the numerical viscous flux contribution is:

\begin{equation}\label{eqn:visc_derivative}
    \left(\frac{\partial \widetilde{\mathbf{F}}^v}{\partial x}\right)_{j} = \frac{\widetilde{\mathbf{F}}^v_{j+\frac{1}{2}} - \widetilde{\mathbf{F}}^v_{j-\frac{1}{2}}}{\Delta x},
\end{equation}

\noindent where $\widetilde{(\cdot)}$ denotes the \textit{numerical} flux, which is distinct from the physical flux. The numerical viscous flux at the interface can be computed as:

\begin{equation}
    \widetilde{\mathbf{F}}^v_{j+\frac{1}{2}} = 
    \begin{pmatrix}
        0 \\
        -\tau_{j+\frac{1}{2}} \\
        -\tau_{j+\frac{1}{2}} u_{j+\frac{1}{2}} + q_{j+\frac{1}{2}} \\
    \end{pmatrix},
\end{equation}

\noindent where,

\begin{subequations} \label{eqn:viscInterface}
    \begin{gather}
        \tau_{j+\frac{1}{2}} = \frac{4}{3} \hat{\mu}_{j+\frac{1}{2}} \left( \frac{\partial u}{\partial x} \right)_{j+\frac{1}{2}},
        \quad
        q_{j+1 / 2} = -\frac{\hat{\mu}_{j+\frac{1}{2}}}{(\gamma-1) \mathrm{Pr}}  \left( \frac{\partial T}{\partial x} \right)_{j+\frac{1}{2}}.
        \tag{\theequation a-\theequation b}
    \end{gather}
\end{subequations}

\noindent The non-dimensional dynamic viscosity is found from Sutherland's Law:

\begin{equation}
    \mu_{j+\frac{1}{2}} = T^{3/2}_{j+\frac{1}{2}} \frac{1 + C/T_{\infty}}{T_{j+\frac{1}{2}} + C/T_{\infty}},
\end{equation}

\noindent where $C = \SI{110.4}{\kelvin}$ is Sutherland's constant. \textcolor{black}{For all the $\alpha$-damping schemes (sections \ref{sec_OC6}, \ref{sec_E6}, \ref{sec_E2}) considered in this paper and the Shen's sixth-order scheme (section \ref{shen_visc})}  \textcolor{black}{dynamic viscosity is computed from Sutherland's Law from the temperature at the half locations. On the other hand, for the schemes to be presented in the sections \ref{nad_e6} and \ref{nad_oc6}, the dynamic viscosity at the cell center is computed from the temperature at the cell center itself as in the original references.} Quantities at half locations are computed via an arithmetic average of the left and right reconstructed states. The reconstructed states are:

\begin{subequations} \label{uLuR_damp_sixth}
    \begin{gather}
        u_L = u_j + \left(   \frac{ \partial u }{  \partial x }   \right)_{j} \frac{\Delta x}{2} + \beta  \left({ {u}}_{j+1} - 2{ {u}}_{j} + { {u}}_{j-1}\right),
        \tag{\theequation a} \\
        u_R =  u_{j+1}  - \left(   \frac{ \partial u }{  \partial x }   \right)_{j+1} \frac{\Delta x}{2} + \beta  \left({ {u}}_{j+2} - 2{ {u}}_{j+1} + { {u}}_{j}\right).
        \tag{\theequation b}
    \end{gather}
\end{subequations}

\noindent The $\alpha$-damping approach is then used to approximate the derivatives at the cell-interfaces via:

\begin{equation} \label{good_6tth-order_scheme_flux}
\begin{aligned}
  \left(\frac{\partial u}{\partial x}\right)_{j+ \frac{1}{2}}=\frac{1}{2}\left[\left(\frac{\partial u}{\partial x}\right)_{j}+\left(\frac{\partial u}{\partial x}\right)_{j+1}\right]+\frac{\alpha}{ 2 \Delta x}\left(u_R- u_L\right), \\
\end{aligned}
\end{equation}

\noindent where the gradients at the cell-centers are computed using a finite-difference approximation. The terms $\alpha$ and $\beta$ are free parameters, which, in part, determine the order of accuracy of the final viscous flux discretization. Moreover, using different finite difference approximations for these cell-center gradients distinguishes one $\alpha$-damping approach from another.

\subsubsection{$\alpha$-OC6}\label{sec_OC6}

To compute the present $\alpha$-damping approach, we consider OC6 \cite{kim1996optimized} for the cell-center derivatives:

\begin{equation}\label{eqn-OC6}
    \Theta \left( \frac{\partial u}{\partial x} \right)_{j-1} + \left( \frac{\partial u}{\partial x} \right)_{j} + \Theta \left( \frac{\partial u}{\partial x} \right)_{j+1} = a \frac{u_{j+1} - u_{j-1}}{2} + b \frac{u_{j+2} - u_{j-2}}{4} + c \frac{u_{j+3} - u_{j-3}}{6},
\end{equation}

\noindent with $\Theta = \frac{30000}{73425}$, $a = \frac{\Theta + 9}{6}$, $b = \frac{32 \Theta - 9}{15}$, and $c = \frac{-3 \Theta + 1}{10}$. To determine the free parameters $\alpha$ and $\beta$, we perform a Fourier analysis for a scalar diffusion equation with a unit diffusion coefficient, as described by Chamarthi et al. \cite{chamarthi2022}. The use of OC6 for the gradients in the $\alpha$-damping viscous flux discretization is termed $\alpha$-OC6. To achieve sixth-order accurate viscous fluxes, we set $\alpha = \frac{14}{3}$ and $\beta = \frac{1}{84}$. Accordingly, the modified wavenumber for $\alpha$-OC6 was found to be:

\begin{equation}
\mathcal{F}(k)_{\alpha-OC6} = -\frac{2 \sin ^2\left(\frac{k}{2}\right) (61713 \cos (k)-5094 \cos (2 k)+442 \cos (3 k)+103049)}{45 (800 \cos (k)+979)}	,
\end{equation}

\noindent where $k$ is the wavenumber. This can be expanded to:

\begin{equation}
\mathcal{F}(k)_{\alpha - OC6} = -k^2 \left(1-\frac{5987 k^6}{8966160}+\frac{1782961 k^{8}}{11393427600}+O\left(k^{9}\right)	\right),
\end{equation}

\noindent which clearly shows the sixth- and higher-order error terms. Fig. \ref{fig:second_deriv} displays the comparison of second derivative differencing error for $\alpha$-OC6 and the theoretical curve.

\subsubsection{$\alpha$-E6}\label{sec_E6}

To compare, we also consider the sixth-order viscous fluxes derived in \cite{chamarthi2022}, denoted as $\alpha$-E6 in this paper. This viscous flux discretization uses an explicit fourth-order finite difference scheme for the gradients and free parameters $\alpha = \frac{38}{15}$, $\beta = -\frac{11}{228}$. $\alpha$-E6 has the following modified wavenumber:

\begin{equation}
\mathcal{F}(k)_{\alpha-E6} = -\frac{1}{6} \sin ^2\left(\frac{k}{2}\right) \left(\frac{148}{5} - \frac{92}{15}\cos(k) + \frac{8}{15} \cos(2 k)\right),
\end{equation}

\noindent with the corresponding series expansion,

\begin{equation}
    \mathcal{F}(k)_{\alpha-E6} = -k^2 \left(1 - \frac{k^6}{560} + \frac{k^8}{3600} + O\left(k^{9}\right) \right).
\end{equation}

\noindent Even though the $\alpha$-OC6 and $\alpha$-E6 viscous flux discretizations are both sixth-order accurate, the difference in spectral properties, specifically in the high wavenumber region, is significant, as shown in Fig. \ref{fig:second_deriv}. It will be shown later that this difference helps damp high frequency errors. 

\subsubsection{E2}\label{sec_E2}

Apart from the sixth-order $\alpha$-damping methods, we also considered a standard second-order central viscous flux discretization, denoted as E2 in this paper. E2 can also be considered as an $\alpha$-damping approach (readers may refer to \cite{nishikawa:AIAA2010}). The first derivative at the cell-interfaces for the E2 scheme is computed as follows:

\begin{subequations}
    \begin{gather}
        \left( \frac{\partial u}{\partial x} \right)_{j+\frac{1}{2}}=\frac{1}{\Delta x}\left( u_{j+1}-u_{j}\right), 
        \quad
        \left(\frac{\partial u}{\partial x}\right)_{j-\frac{1}{2}}=\frac{1}{\Delta x}\left(u_{j}- u_{j-1}\right).
        \tag{\theequation a-\theequation b}
    \end{gather}
\end{subequations}

\noindent The second derivative differencing error of E2 is also shown in Fig. \ref{fig:second_deriv}.

\subsection{Non-$\alpha$-damping (NAD) Approach}

\subsubsection{NAD-E6}\label{nad_e6}

The first NAD approach we present is the sixth-order viscous flux-discretization of Luo et al. \cite{luo2013comparison}. In this method, the cell-center gradients are computed using a sixth-order central-difference scheme. For example, the first derivative of the streamwise velocity is computed as:

\begin{equation}\label{eqn:4e}
    \left( \frac{\partial u}{\partial x} \right)_j = \frac{1}{\Delta x} \left( -\frac{1}{60} u_{j-3} + \frac{3}{20} u_{j-2} - \frac{3}{4} u_{j-1} + \frac{3}{4} u_{j+1} - \frac{3}{20} u_{j+2} + \frac{1}{60} u_{j+3} \right).
\end{equation}

\noindent Once the gradients of the necessary variables are computed, the \textit{cell-center} numerical viscous fluxes, $\widetilde{\mathbf{F}}^v_j$, can be obtained via Equation (\ref{eqn:viscFluxes}a). $\widetilde{\mathbf{F}}^v_j$ are then interpolated to the cell-interfaces by using the sixth-order reconstruction formula:

\begin{equation} \label{sixth}
\widetilde{\mathbf{F}}^v_{j+ \frac{1}{2}} = \frac{1}{60}  \mathbf{\widetilde{F}}^v_{j-2} - \frac{8}{60} \mathbf{\widetilde{F}}^v_{j-1} + \frac{37}{60} \mathbf{\widetilde{F}}^v_{j} +
\frac{37}{60} \mathbf{\widetilde{F}}^v_{j+1} - \frac{8}{60} \mathbf{\widetilde{F}}^v_{j+2} + \frac{1}{60} \mathbf{\widetilde{F}}^v_{j+3}. 
\end{equation}

\noindent Once the cell-interface numerical viscous fluxes, $\widetilde{\mathbf{F}}^v_{j+ \frac{1}{2}}$, are obtained, the viscous flux derivatives are computed via Equation (\ref{eqn:visc_derivative}). We denote this scheme as NAD-E6. Fourier analysis of this scheme is carried out and the modified wavenumber can be expressed as: 

\begin{equation}
\mathcal{F}(k)_{NAD-E6}= -\frac{1}{900} (45 \sin (k)-9 \sin (2 k)+\sin (3 k))^2,
\end{equation}

\noindent with the corresponding series expansion,

\begin{equation}
\mathcal{F}(k)_{NAD-E6} = -k^2 \left(1 - \frac{k^6}{70}+\frac{k^{8}}{360}+O\left(k^{9}\right)	\right),
\end{equation}

\noindent which clearly shows the sixth- and higher-order error terms. \\

\subsubsection{NAD-OC6} \label{nad_oc6}

The second approach we consider is that of Visbal and Gaitonde \cite{visbal2002use}. In this method, the viscous flux derivatives are computed using an optimized sixth-order compact scheme in lieu of Equation (\ref{eqn:visc_derivative}). First, the velocity and temperature gradients are computed with OC6. Then, $\widetilde{\mathbf{F}}^v_j$ are computed from Equation (\ref{eqn:viscFluxes}a). Then, the viscous flux derivatives are computed from:

\begin{equation}
    \Theta    
    \left(\frac{\partial \widetilde{\mathbf{F}}^v}{\partial x} \right)_{j-1} + \left(\frac{\partial \widetilde{\mathbf{F}}^v}{\partial x} \right)_{j} + \Theta \left(\frac{\partial \widetilde{\mathbf{F}}^v}{\partial x} \right)_{j+1} = a \frac{ \widetilde{\mathbf{F}}^v_{j+1} - \widetilde{\mathbf{F}}^v_{j-1}}{2} + b \frac{\widetilde{\mathbf{F}}^v_{j+2} - \widetilde{\mathbf{F}}^v_{j-2}}{4} + c \frac{\widetilde{\mathbf{F}}^v_{j+3} - \widetilde{\mathbf{F}}^v_{j-3}}{6},
\end{equation}

\noindent with $\Theta = \frac{30000}{73425}$, $a = \frac{\Theta + 9}{6}$, $b = \frac{32 \Theta - 9}{15}$, and $c = \frac{-3 \Theta + 1}{10}$. This approach is denoted NAD-OC6. Fourier analysis of this approach is carried out and the modified wavenumber is:

\begin{equation}
\mathcal{F}(k)_{NAD-OC6} = -\frac{ \sin ^2(k) (3989 \cos (k)-221 \cos (2 k)+22917)^2}{225 (800 \cos (k)+979)^2},
\end{equation}

\noindent where the above equation can be expanded as follows,

\begin{equation}
\mathcal{F}(k)_{NAD-OC6} = -k^2\left(1+\frac{263 k^6}{373590}-\frac{4117 k^{8}}{42197880}-O\left(k^{9}\right)\right).
\end{equation}

\noindent It can be seen that the NAD-OC6 approach is also sixth-order accurate. \\

\subsubsection{NAD-Shen}\label{shen_visc}

The third NAD method we considered was the sixth-order conservative difference method of Shen et al. \cite{shen2010large}, denoted as NAD-Shen. The Fourier analysis of NAD-Shen is carried out in \cite{chamarthi2022} and is not repeated here. 

\subsection{\textcolor{black}{Spectral Properties of the Considered Viscous Flux Discretizations}}

\begin{figure}[h!]
    \centering
    \includegraphics[width=0.5\textwidth]{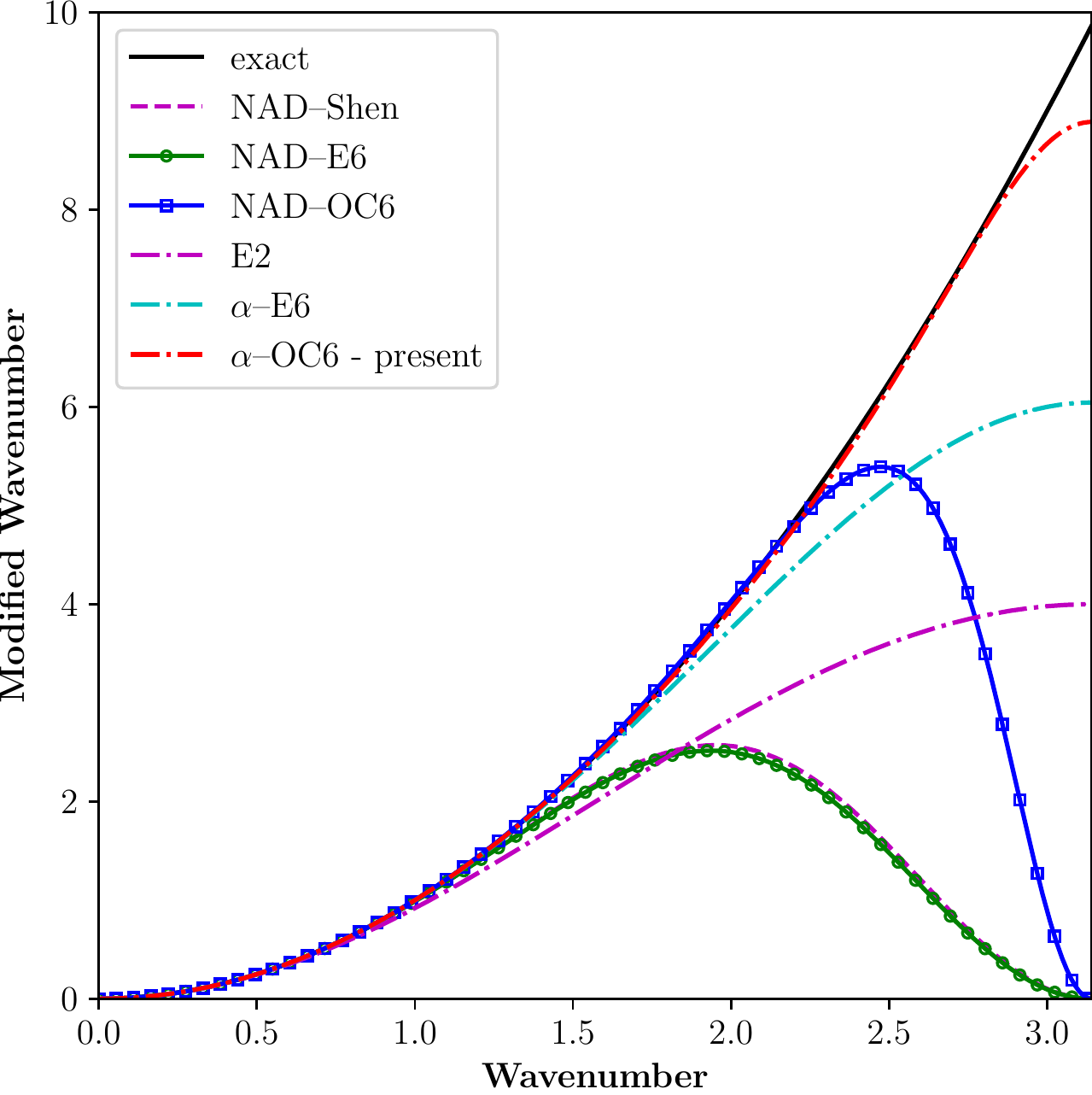}
    \caption{Differencing error for second derivative vs. wavenumber.}
    \label{fig:second_deriv}
\end{figure}

Figure \ref{fig:second_deriv} displays the differencing error for the second derivative for the considered methods. The results were obtained by performing a Fourier analysis for a scalar diffusion equation with a unit diffusion coefficient. It is evident that the $\alpha$-OC6 method matches very closely with the theoretical curve and thus better represents the physical dissipation at high wavenumbers. While the $\alpha$-E6 and E2 methods perform better over a larger range of wavenumbers than the NAD approaches, they become inaccurate towards higher wavenumbers. Furthermore, while the NAD-Shen, NAD-E6, and NAD-OC6 methods are all sixth-order accurate viscous flux discretizations, they have poor spectral properties in the high wavenumber range relative to the $\alpha$-damping methods. One reason for this is that these NAD methods compute the second derivative by applying the first derivatives twice, which according to Nagarajan et al. \cite{nagarajan2003robust} will lead to poor spectral properties. A secondary observation here is that there is negligible difference between Shen et al.'s \cite{shen2010large} approach and that of Luo et al. \cite{luo2013comparison} in terms of spectral properties.

\subsection{Time Integration}

Eq. (\ref{eqn:cns}) is cast in semi-discrete form to allow for temporal integration of the right-hand-side. The explicit third-order total-variation-diminishing Runge-Kutta (RK3TVD) \cite{Jiang1995} method is used for time integration. The timestep, $\Delta t$, was computed from the CFL condition. We used both an inviscid and a viscous analogue of the CFL condition. For all simulations, $\mathrm{CFL} = 0.1$ to better ensure numerical stability. The inviscid $\Delta t$ was computed from:

\begin{equation}
    \Delta t^c = \min \left( \frac{\Delta x}{\lvert u \rvert + a}, \frac{\Delta y}{\lvert v \rvert + a}, \frac{\Delta z}{\lvert w \rvert + a} \right), 
\end{equation}

\noindent where $a = \sqrt{\gamma p/\rho}$ is the local speed of sound. The viscous $\Delta t$ was computed from:

\begin{equation}
    \Delta t^v = \frac{1}{\alpha} \min \left( \frac{\Delta x^2}{\hat{\nu}}, \frac{\Delta y^2}{\hat{\nu}}, \frac{\Delta z^2}{\hat{\nu}} \right),
\end{equation}

\noindent where $\alpha$ corresponds to that employed in the viscous spatial discretization method and $\hat{\nu} = \hat{\mu}/\rho$ is the local, scaled kinematic viscosity. Finally, the current timestep's $\Delta t$ was:

\begin{equation}
    \Delta t = \mathrm{CFL} \times \min \left( \Delta t^c, \Delta t^v \right).
\end{equation}

\section{Results}\label{sec4}

\subsection{Viscous TGV}\label{ex:vTGV}

The viscous TGV is a classic test case used to assess a scheme's ability to resolve the flowfield's inviscid and viscous features. The initial conditions are as follows:

\begin{equation}\label{vtgv}
    \begin{pmatrix}
        u \\
        v \\
        w \\
        p \\
        T
    \end{pmatrix}
    =
    \begin{pmatrix}
         \sin(x) \cos (y) \sin(z) \\
        - \cos (x) \sin(y) \sin(z) \\
        0 \\
        \frac{1}{\gamma M^2}+\frac{1}{16}\left(\cos(2x)+\cos(2y)\right)\left(\cos(2z)+2\right) \\
         1
    \end{pmatrix}.
\end{equation}

Simulations were carried out for three Reynolds numbers: $\mathrm{Re} = 800$, $\mathrm{Re} = 1600$, and $\mathrm{Re} = 3000$, in a periodic domain of $[0,2\pi]^3$ all with $\gamma = 5/3$. To ensure incompressibility, $\mathrm{Ma} = 0.1$. In all simulations, Sutherland's law was used to compute the dynamic viscosity with reference temperature $T_{\infty} = \SI{300}{\kelvin}$. The flow was simulated at under-resolved grid resolutions of $64^3$, $96^3$, and $128^3$ for $\mathrm{Re}= 800$, $\mathrm{Re}= 1600$, and $\mathrm{Re}= 3000$, respectively. The simulations were performed until an end time of $t = 10$. As the simulation progressed, the temporal evolution of enstrophy was monitored to compare with the available DNS data \cite{brachet1983small}. \textcolor{black}{Simulations using sixth-order KEEP scheme, second-order KEEP scheme and ninth-order upwind scheme are presented below.}\\

\begin{figure}[h!]
\centering
\subfigure[$\mathrm{Re}=800$ for NAD methods.]{\includegraphics[width=0.48\textwidth]{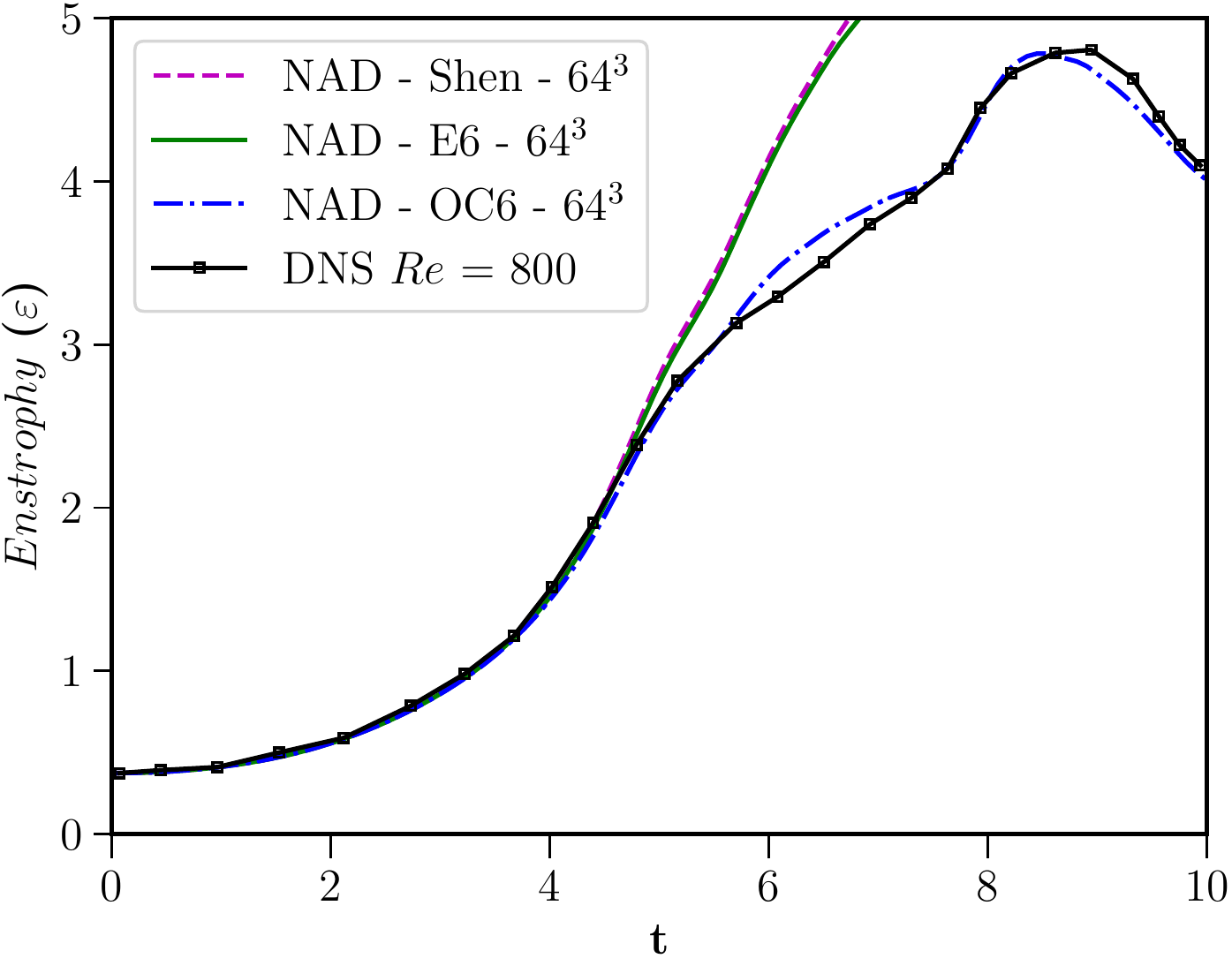}
\label{fig:bad1}}
\subfigure[$\mathrm{Re}=800$ for $\alpha$-damping methods.]{\includegraphics[width=0.48\textwidth]{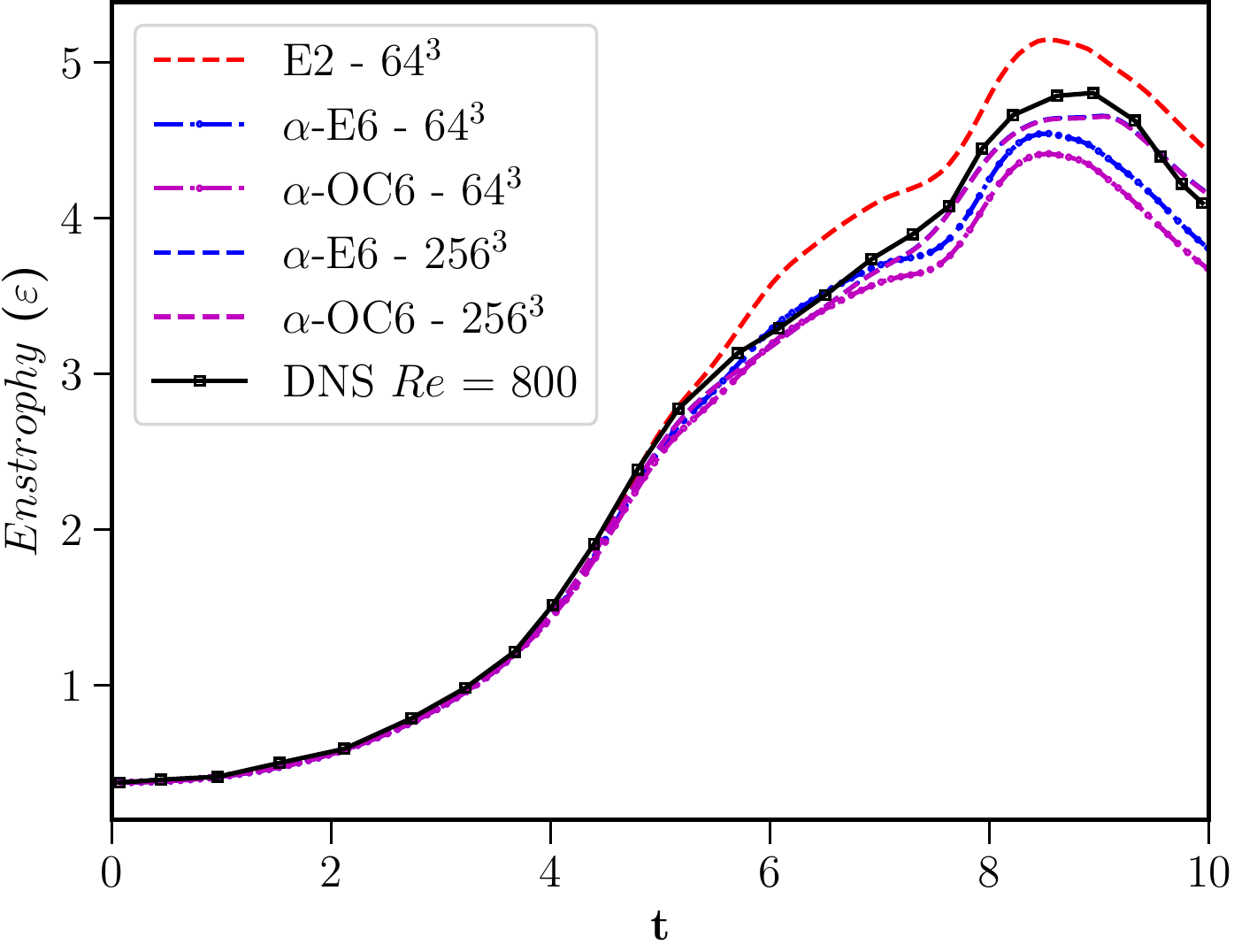}
\label{fig:good1}}
\caption{Enstrophy of viscous TGV for $\mathrm{Re}=800$ with various grid resolutions and methods.}
\label{fig_vTGV_800}
\end{figure}

\begin{figure}[h!]
\centering
\subfigure[$\mathrm{Re}=3000$ for NAD methods.]{\includegraphics[width=0.48\textwidth]{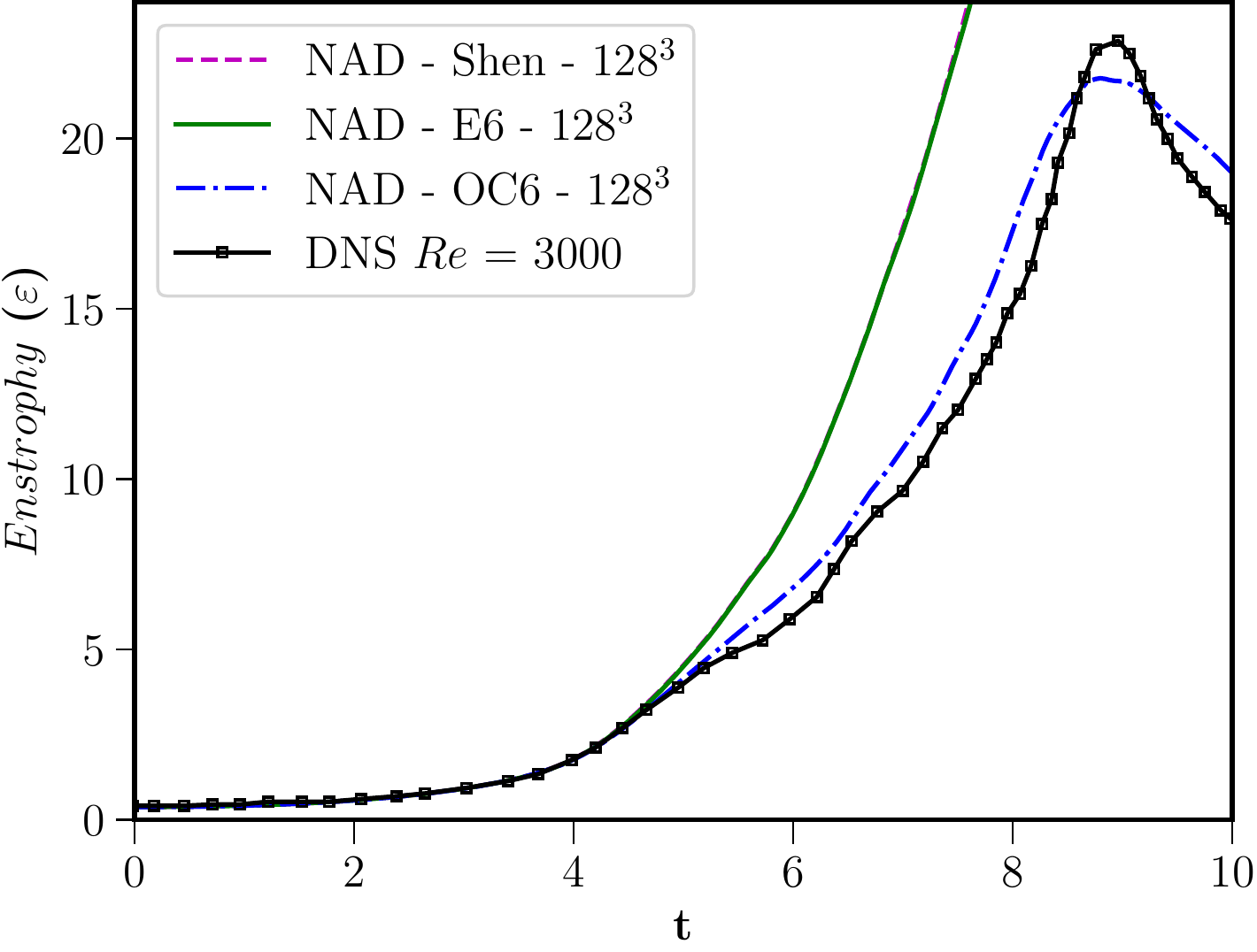}
\label{fig:bad2}}
\subfigure[$\mathrm{Re}=3000$ for $\alpha$-damping methods.]{\includegraphics[width=0.48\textwidth]{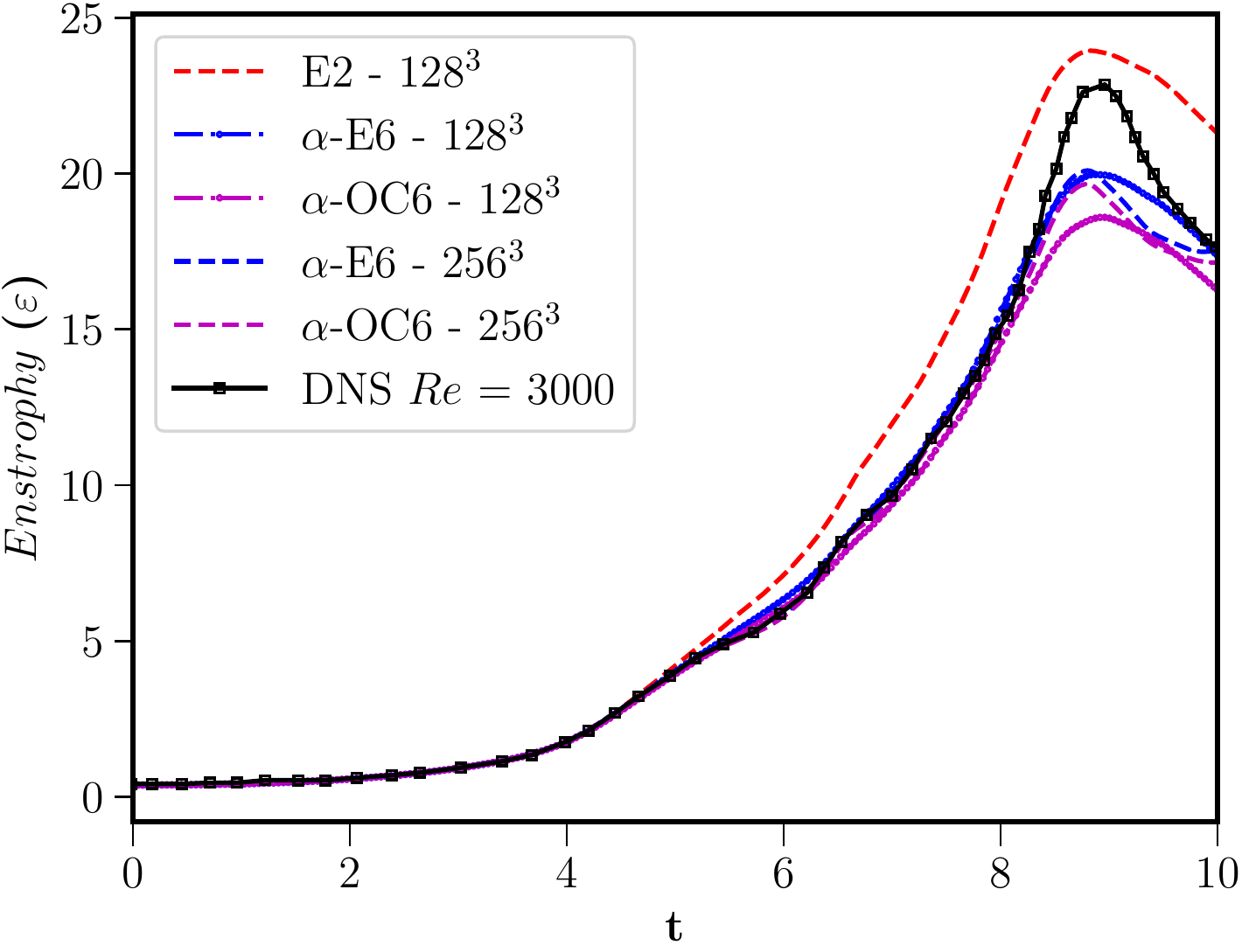}
\label{fig:good3}}
\caption{Enstrophy of viscous TGV for $\mathrm{Re}=3000$ with various grid resolutions and methods.}
\label{fig_vTGV_3000}
\end{figure}

\begin{figure}[H]
\centering
\subfigure[$\mathrm{Re}=1600$ for NAD methods.]{\includegraphics[width=0.48\textwidth]{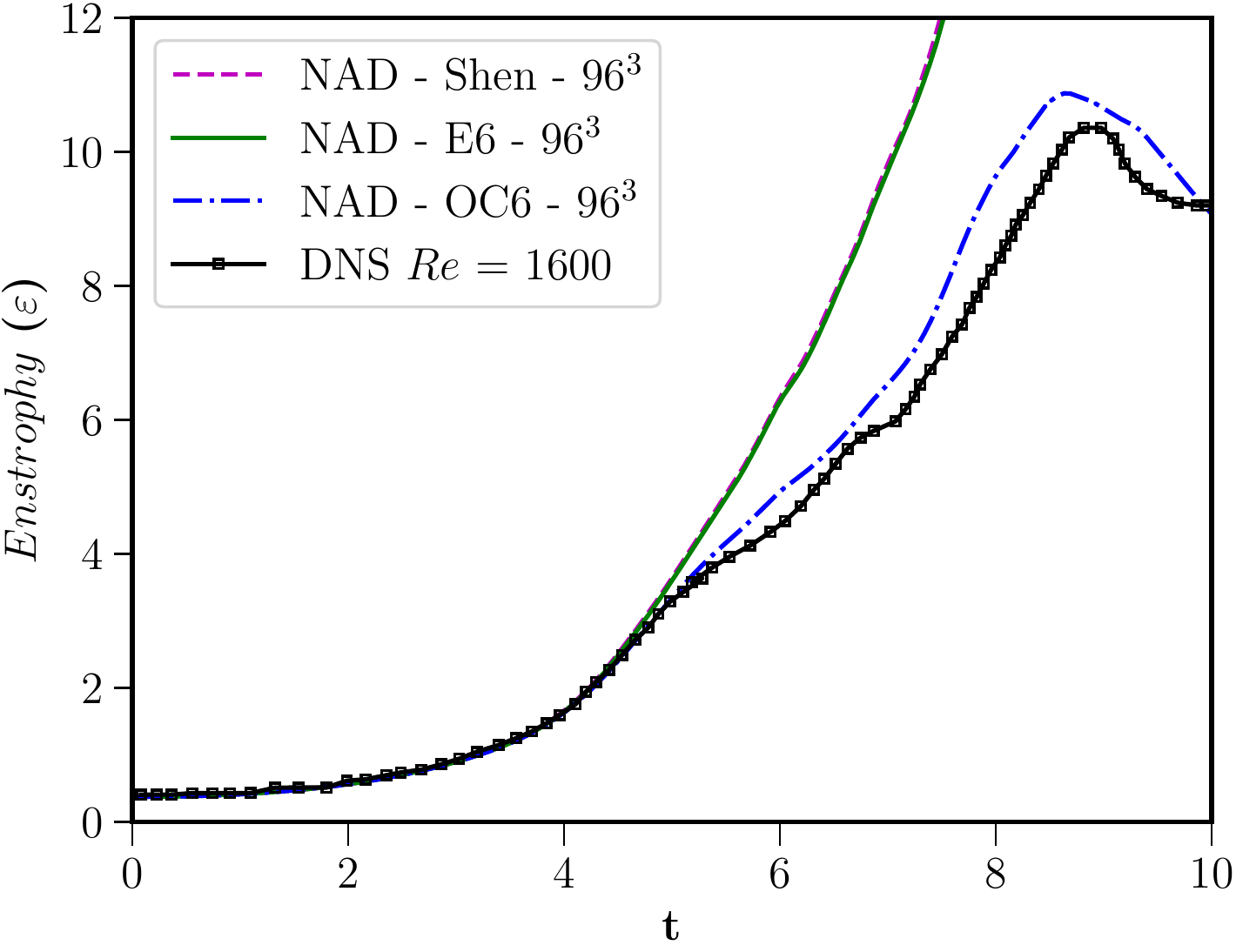}
\label{fig:bad3}}
\subfigure[$\mathrm{Re}=1600$ from Ref. \cite{abe2018stable}]{\includegraphics[width=0.42\textwidth]{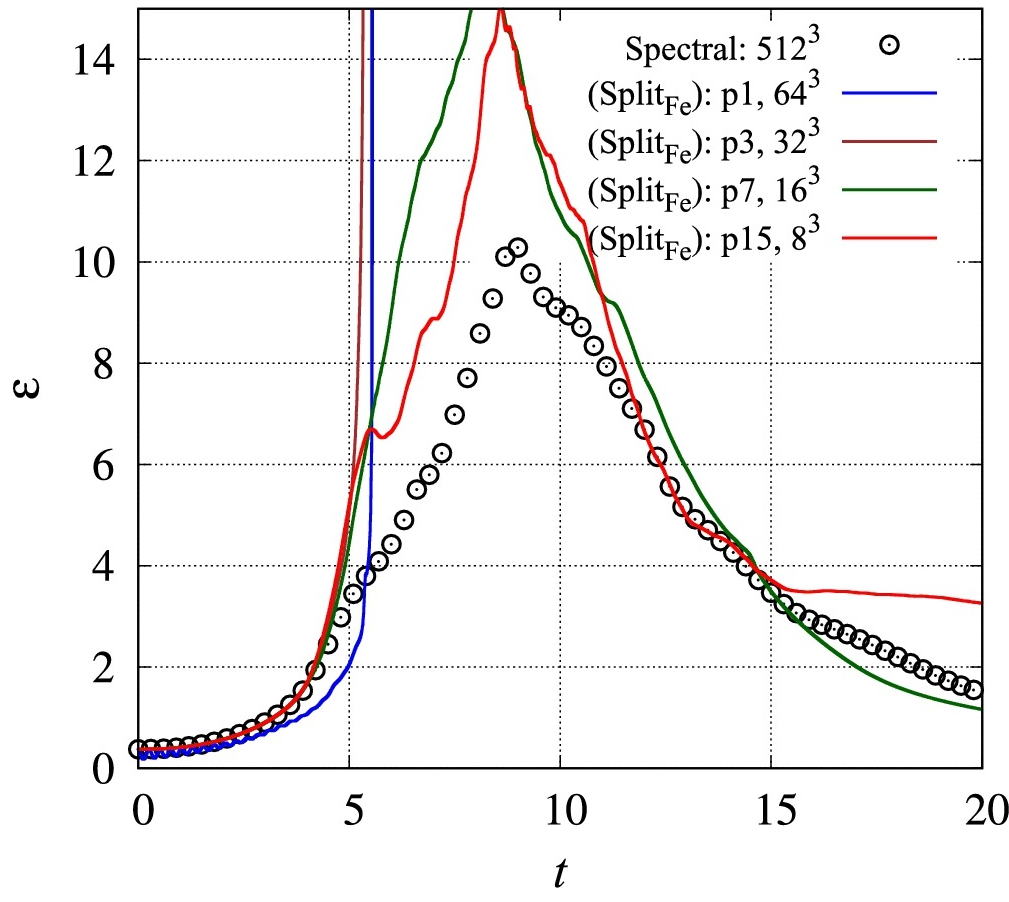}
\label{fig:bad_dg}}
\caption{Fig. \ref{fig:bad3} Left: Enstrophy of viscous TGV for $\mathrm{Re}=1600$ with various grid resolutions using NAD methods. Right: taken from Ref. \cite{abe2018stable}.}
\label{fig_vTGV_1600_bad}
\end{figure}

\begin{figure}[H]
\centering
\subfigure[$\mathrm{Re}=1600$ for $\alpha$-damping methods.]{\includegraphics[width=0.48\textwidth]{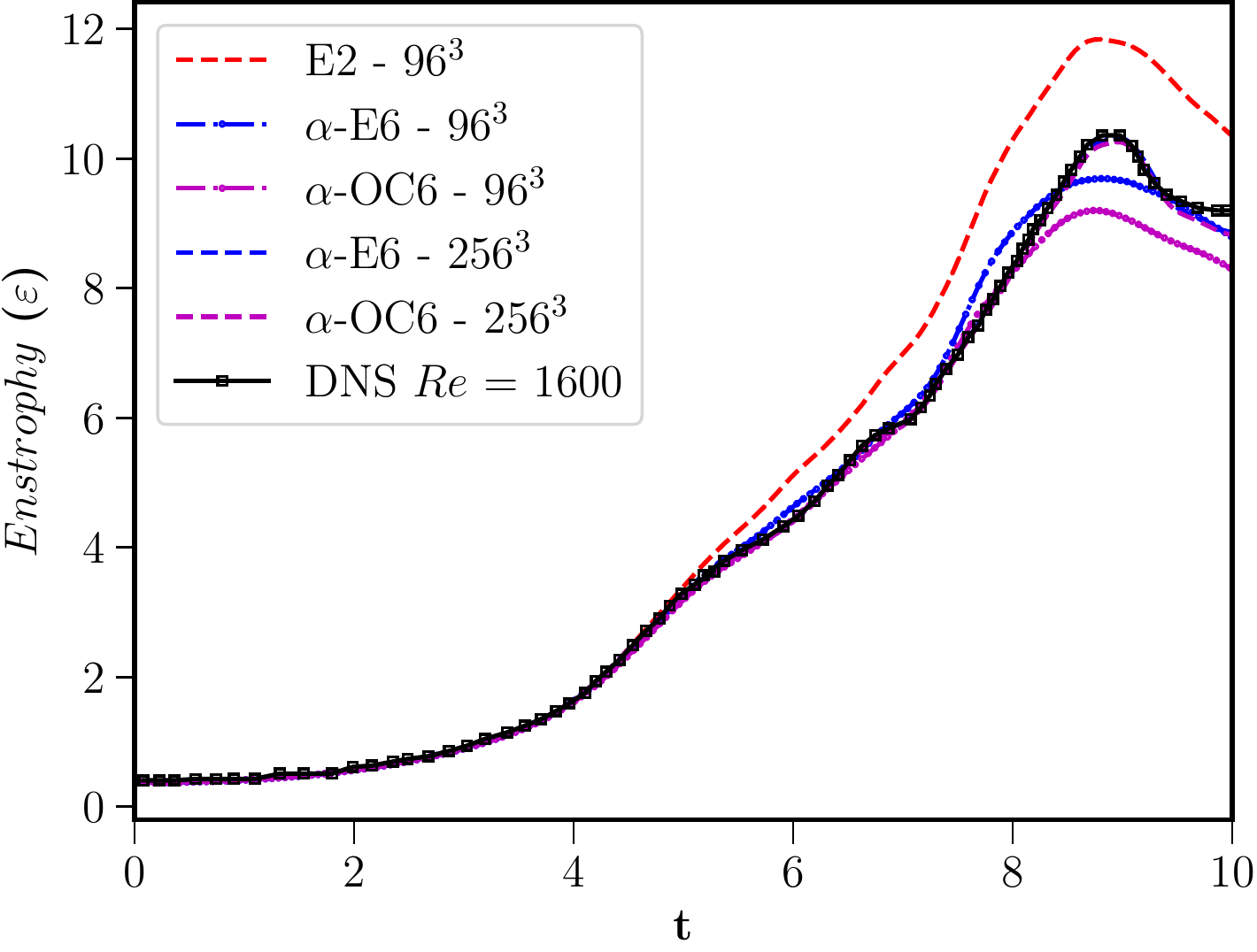}
\label{fig:good2}}
\subfigure[$\mathrm{Re}=1600$ from Ref. \cite{abe2018stable} ]{\includegraphics[width=0.48\textwidth]{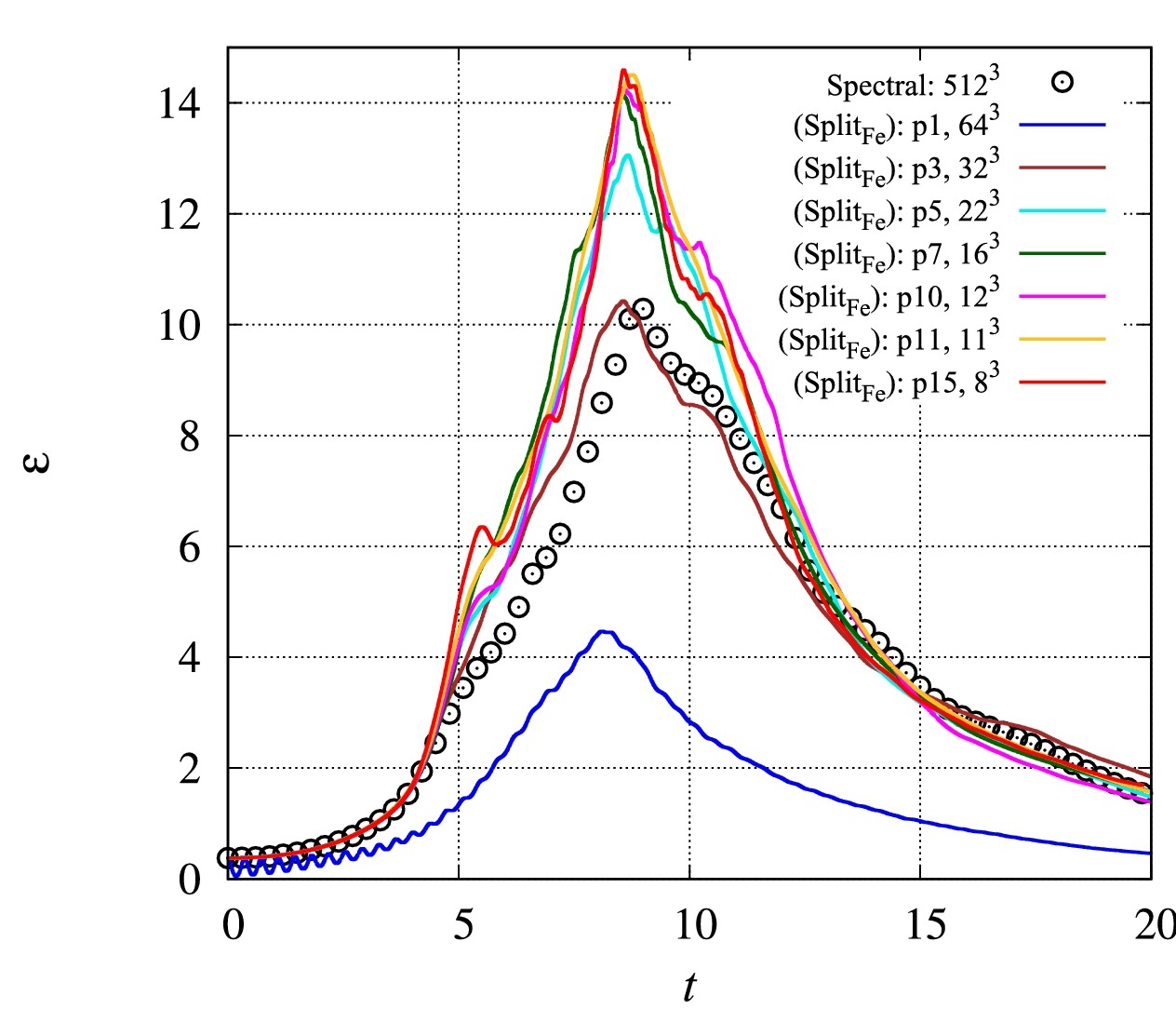}
\label{fig:stillbad}}
\caption{Left: Enstrophy of viscous TGV for $\mathrm{Re}=1600$ with various grid resolutions using $\alpha$-damping schemes. Right: taken from Ref. \cite{abe2018stable}.}
\label{fig_vTGV_very_1600_good}
\end{figure}

\textcolor{black}{\textbf{Simulations using sixth-order KEEP scheme:}}  A series of plots depicting the temporal evolution of enstrophy for different Reynolds numbers are shown in Figs. \ref{fig_vTGV_800}-\ref{fig_vTGV_very_1600_good}. The following observations can be made:

\begin{itemize}

\item Figs. \ref{fig:bad1} and \ref{fig:good1} show the enstrophy profiles obtained using the NAD and $\alpha$-damping schemes, respectively, for the $\mathrm{Re} = 800$ simulation using a grid size of $64^3$. All NAD schemes over-predicted the enstrophy values compared to the  DNS data carried out on a grid size of $512^3$, which indicates physical inconsistency since this would mean that more turbulence was resolved than the reference DNS on a coarser grid. This inconsistency can be attributed to the spectral properties of these schemes, shown in Fig. \ref{fig:second_deriv}. Nagarajan et al. \cite{nagarajan2003robust} have also observed that the schemes that compute the second derivatives by applying the first derivatives twice do indeed have poor damping characteristics in the high-wavenumber region.

\item Fig. \ref{fig:good1} shows the results obtained by the $\alpha$-damping schemes. $\alpha$-E6 and $\alpha$-OC6 slightly under-predict the DNS data for both the $64^3$ and $256^3$ grid resolutions, which is physically consistent since they are under-resolved. However, similar to the NAD schemes, E2 over-predicts the DNS data due to the insufficient spectral resolution in the high wavenumber region, as shown in Fig. \ref{fig:second_deriv}. Consistent with the spectral properties of the concerned schemes, $\alpha$-OC6 provides damping over a larger wavenumber range, which results in a slightly higher measure of enstrophy for $\alpha$-E6.

\end{itemize}

\begin{itemize}

\item Furthermore, as shown in Figs \ref{fig:bad2} and \ref{fig:good3}, similar observations can be made for $\mathrm{Re}=3000$. The enstrophy value for $\mathrm{Re}=3000$ using the NAD schemes is well beyond the DNS results on a coarse grid of $128^3$ while the results obtained for $\alpha$-damping schemes, $\alpha$-E6 and $\alpha$-OC6, are well within the DNS data for all the grids. The E2 scheme once again over-predicted the enstrophy value due to a lack of sufficient damping.

\item Enstrophy values for $\mathrm{Re}=1600$ are shown in Figs. \ref{fig:bad3} and \ref{fig:good2} for the  NAD and $\alpha$-damping schemes, respectively. For $\mathrm{Re}=1600$, Abe et al. \cite{abe2018stable} observed similar trends in the context of Discontinuous Galerkin type methods using the kinetic energy preserving approach, shown in Fig. \ref{fig:bad_dg} which is taken from \cite{abe2018stable} (their Figure 10(c)). Note that they show their results up to $t = 20$. However, up until $t = 10$, the enstrophy plot comparison should be similar. It can be seen that the enstrophy results overshot the DNS results even in their case. Upon observing these results, in the same work, Abe et al. \cite{abe2018stable} then modified the BR2 viscous scheme \cite{bassi1997high} to improve the solution characteristics (see Appendix E). The results obtained by the modified viscous scheme are shown in Fig. \ref{fig:stillbad}. The enstrophy values still overshoot the DNS result even with the modification. These results indicate that the viscous flux discretization used by Abe et al. \cite{abe2018stable} may also lack damping characteristics in the high-wavenumber region. It may be possible to improve their results by using the $\alpha$-damping approach proposed by Nishikawa \cite{nishikawa:AIAA2010} in the context of the Discontinuous Galerkin type methods. However, this is beyond the scope of this paper.

\end{itemize}

\begin{figure}[H]
\centering
\subfigure[\textcolor{black}{$\mathrm{Re}=1600$ for NAD methods.}]{\includegraphics[width=0.48\textwidth]{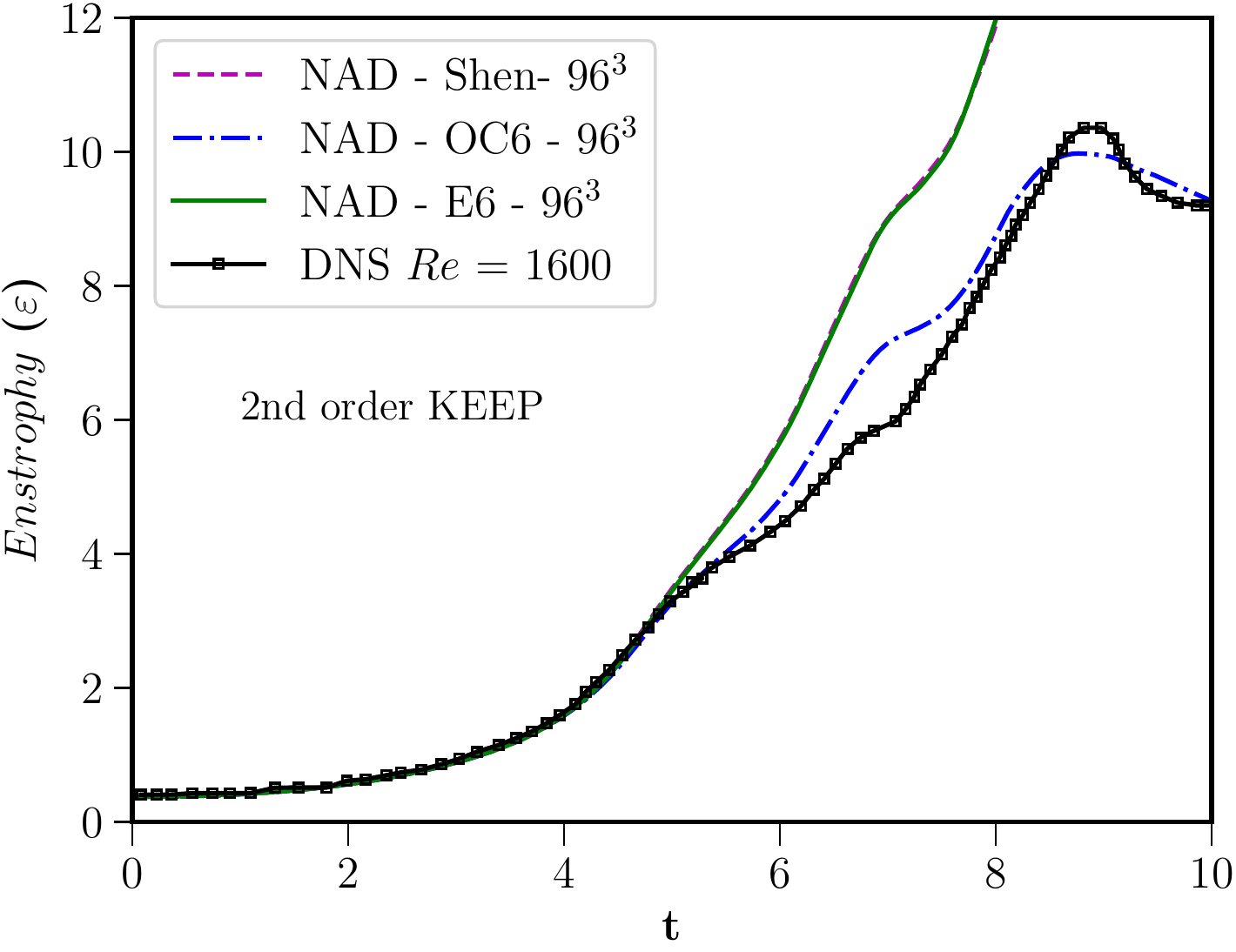}
\label{fig:bad22}}
\subfigure[\textcolor{black}{$\mathrm{Re}=1600$ for $\alpha$-damping methods.}]{\includegraphics[width=0.48\textwidth]{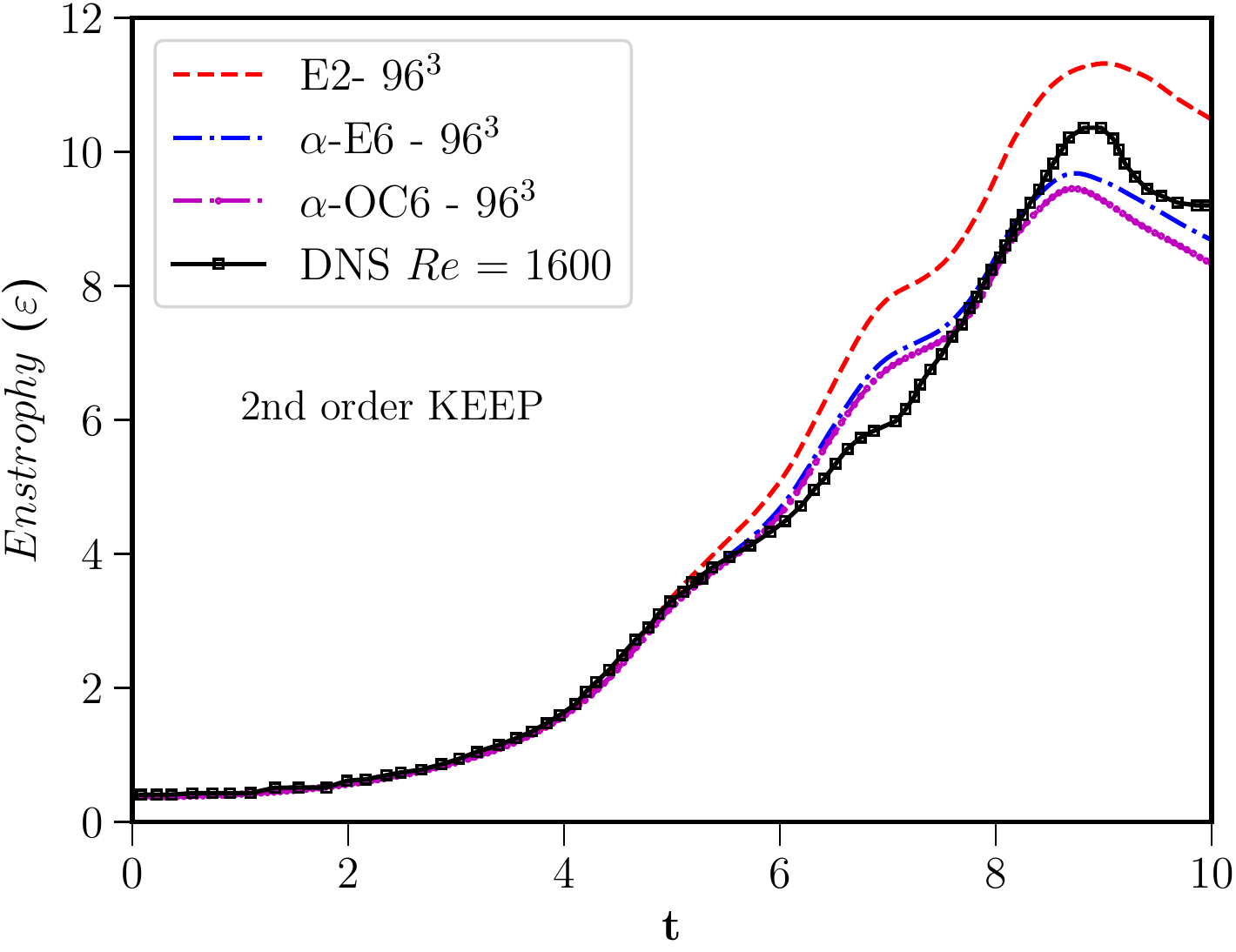}
\label{fig:good32}}
\caption{\textcolor{black}{Enstrophy of viscous TGV for $\mathrm{Re}=1600$ with various grid resolutions and methods using second order KEEP scheme as opposed to sixth-order KEEP considered in earlier figures.}}
\label{fig_2nd_VTGV}
\end{figure}

\textcolor{black}{\textbf{Simulations using second-order KEEP scheme:}} \textcolor{black}{In Figs. \ref{fig_2nd_VTGV}, numerical simulations carried out using the second-order KEEP scheme is presented as opposed to the sixth-order KEEP scheme considered earlier. In Fig. \ref{fig:bad22} NAD schemes are used for the simulations of $\mathrm{Re}=1600$ and the results are similar to that of the Fig. \ref{fig:bad3}. The NAD schemes are over-predicting the enstrophy significantly. On the other hand, the results obtained by the $\alpha$-damping methods are shown in Fig. \ref{fig:good32}. There is some over-prediction of the enstropy between times $t=6$ and $t=8$ as opposed to the sixth-order KEEP scheme. It could be due to the difference in dispersion properties of the second-order and sixth-order inviscid schemes. The overall peak and trend are still under the DNS values. This implies that inviscid scheme also has a little effect on the flow solution (with sixth order discretization performing better than second). However, the effect of viscous is still far more dominating on the flow physics which confirms that the inferences drawn from the sixth order KEEP simulations previously are correct.} 

\textcolor{black}{\textbf{Simulations using ninth-order upwind scheme:}} To demonstrate the importance of using a non-dissipative scheme to isolate the effect of the viscous flux discretization, we ran the inviscid and viscous TGV cases using a linear, upwind, ninth-order reconstruction \cite{Shu1997} for the discretization of the inviscid fluxes in place of solving the KEEP form of the governing equations. For these simulations, we solved the governing equations in divergence form, used the local Lax-Friedrichs flux splitting approach, and considered all aforementioned viscous flux discretization methods. \\

\begin{figure}[H]
\centering
\subfigure[\textcolor{black}{}]{\includegraphics[width=0.48\textwidth]{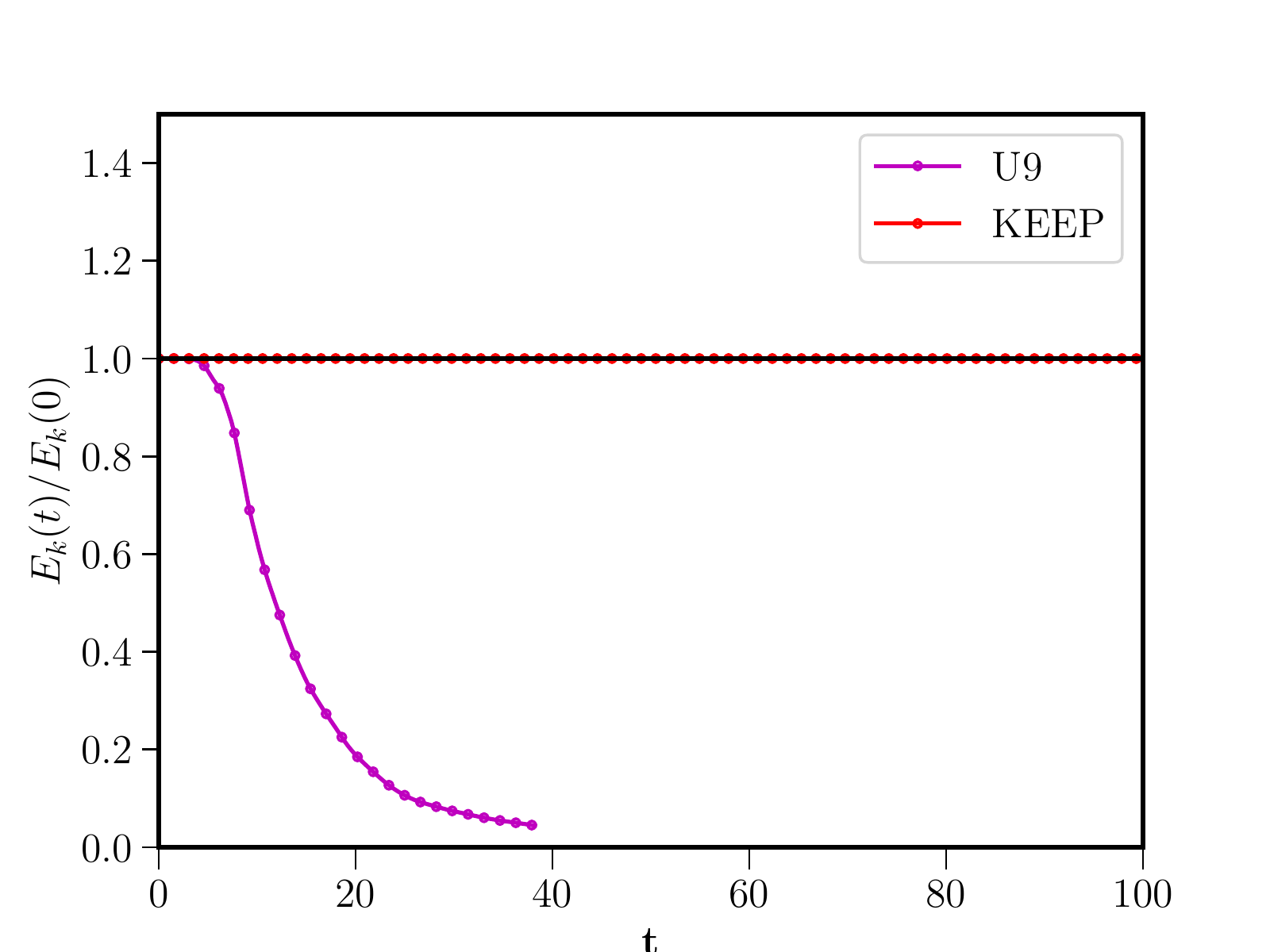}
\label{fig:TGV_U9_KE}}
\subfigure[]{\includegraphics[width=0.48\textwidth]{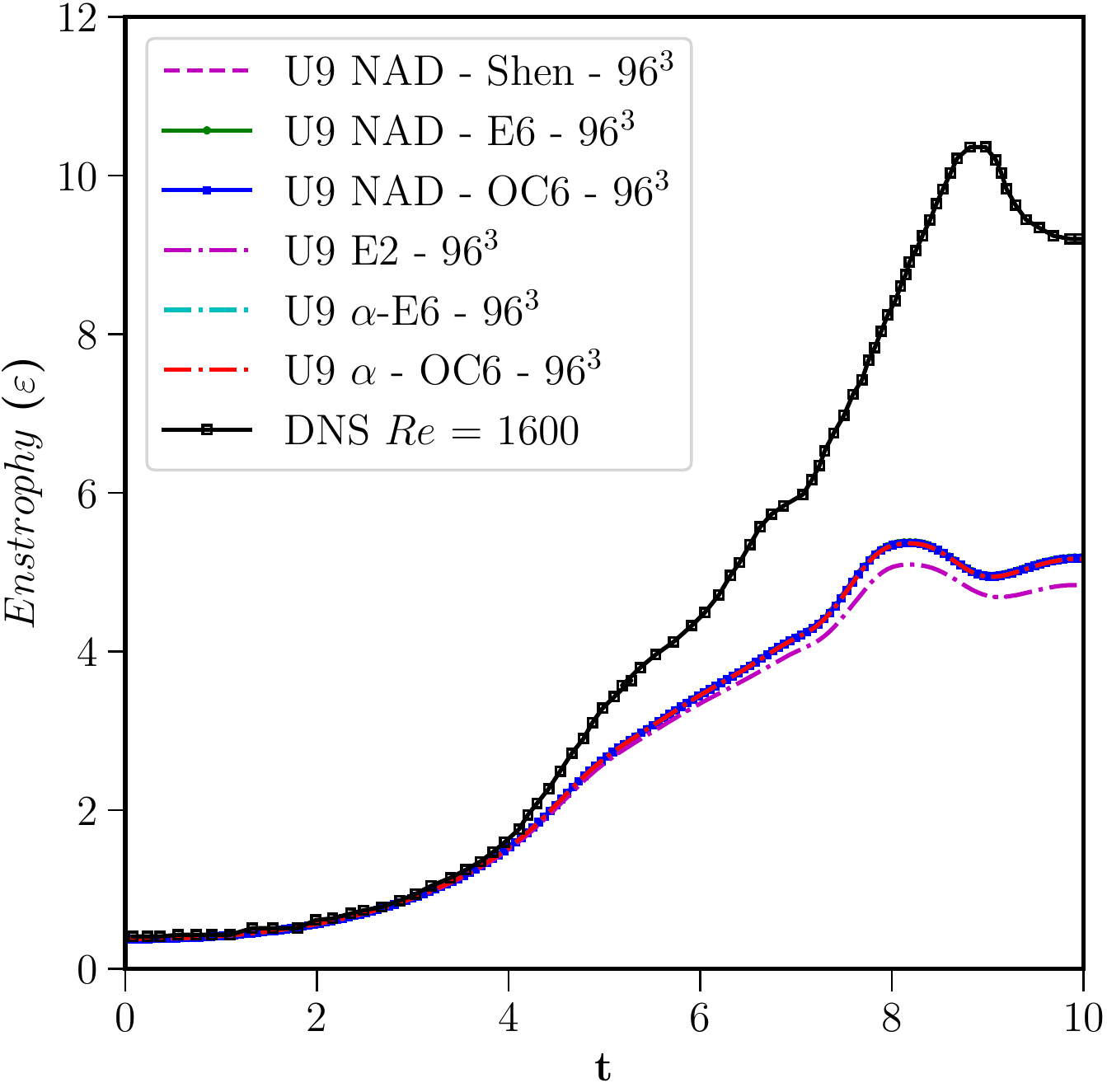}
\label{fig:TGV_ens_u9}}
\caption{\textcolor{black}{Normalized kinetic energy is shown in Fig. (a) and enstrophy for inviscid TGV is shown in Fig. (b). Solid black line: exact solution. Red circles: numerical kinetic energy. Red dashed lines: enstrophy with OC6. Blue solid line: enstrophy with E6.}}
\label{fig_TGV_u9}
\end{figure}

Observing Fig. \ref{fig:TGV_U9_KE}, which shows the volume-averaged kinetic energy computed from the inviscid TGV case, U9 does not preserve kinetic energy. Thus, even for the inviscid case, dissipation exists. Fig. \ref{fig:TGV_ens_u9} displays the computed enstrophy for all viscous flux discretization methods with the U9 scheme applied for the inviscid fluxes. All viscous flux discretizations yield similar enstrophy since the dominant part of the dissipation is from the upwinding of the inviscid fluxes. Therefore, it becomes difficult to monitor the effects of the viscous flux discretization. This may be why it has been previously thought that viscous flux discretization is less important to turbulent flow simulations than inviscid flux discretization. Additionally, it has been previously shown in Ref. \cite{chamarthi2022} that for compressible flows with discontinuities, the NAD-Shen scheme results in odd-even decoupling despite an upwinding approach for inviscid fluxes. While we do not consider flows with discontinuities here, this is important to note, since for many other cases where a dissipative inviscid flux discretization would be applied, a significant error still may result from improper viscous flux discretization. \\

\subsection{Double Periodic Shear Layer}\label{ex:dpsl}

\begin{figure}[H]
    \centering
    \includegraphics[width=0.3\textwidth]{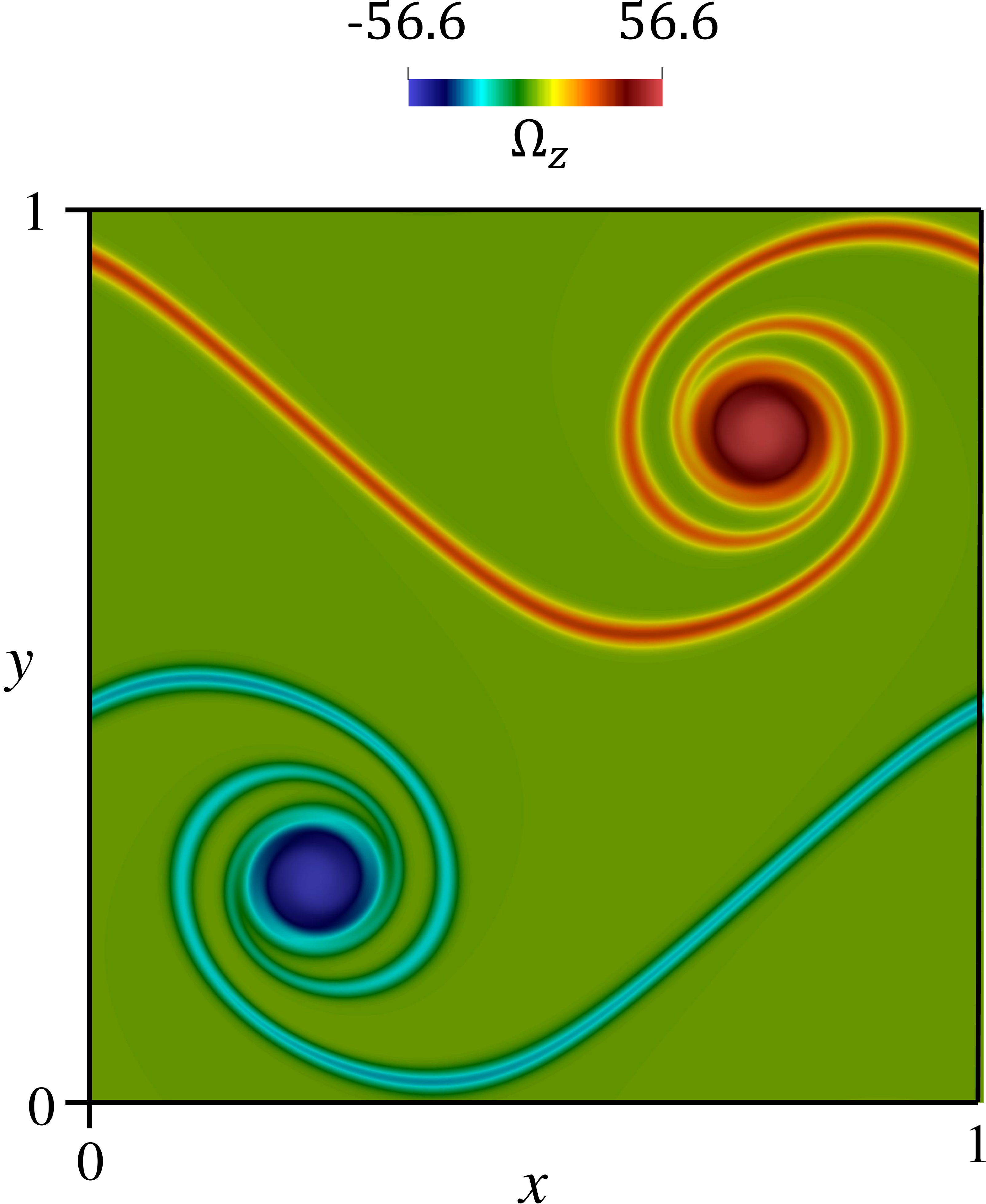}
    \caption{Reference solution - $2048^2$}
    \label{DNS_DPSL}
\end{figure}

This second test case demonstrates how a viscous scheme's spectral properties can affect shear layer resolution. The test case consists of two shear layers initially parallel to each other, which evolve to produce two large vortices at $t=1$. This test case is dominated by viscous forces near the shear layer and is simulated on two grid resolutions: $72^2$ and $144^2$. The non-dimensional parameters are, $\mathrm{Re} = 1 \times 10^4$, $\mathrm{Ma} = 0.1$, and $\gamma = 5/3$. The initial conditions of the flow are:  

\begin{subequations}
    \begin{align}
        \rho &= \frac{1}{\gamma \mathrm{Ma}^2}, \\
        u &= 
        \begin{cases}
            \tanh (80 \times(y-0.25)), & \text{ if } (y \leq 0.5), \\
            \tanh (80 \times(0.75-y)), & \text{ if } (y > 0.5),
        \end{cases} \\
        v &= 0.05 \times \sin (2 \pi(x+0.25)) \\
        T &= 1.
    \end{align}
\end{subequations}

\begin{figure}[H]
\centering
\subfigure[NAD - Shen]{\includegraphics[width=0.25\textwidth]{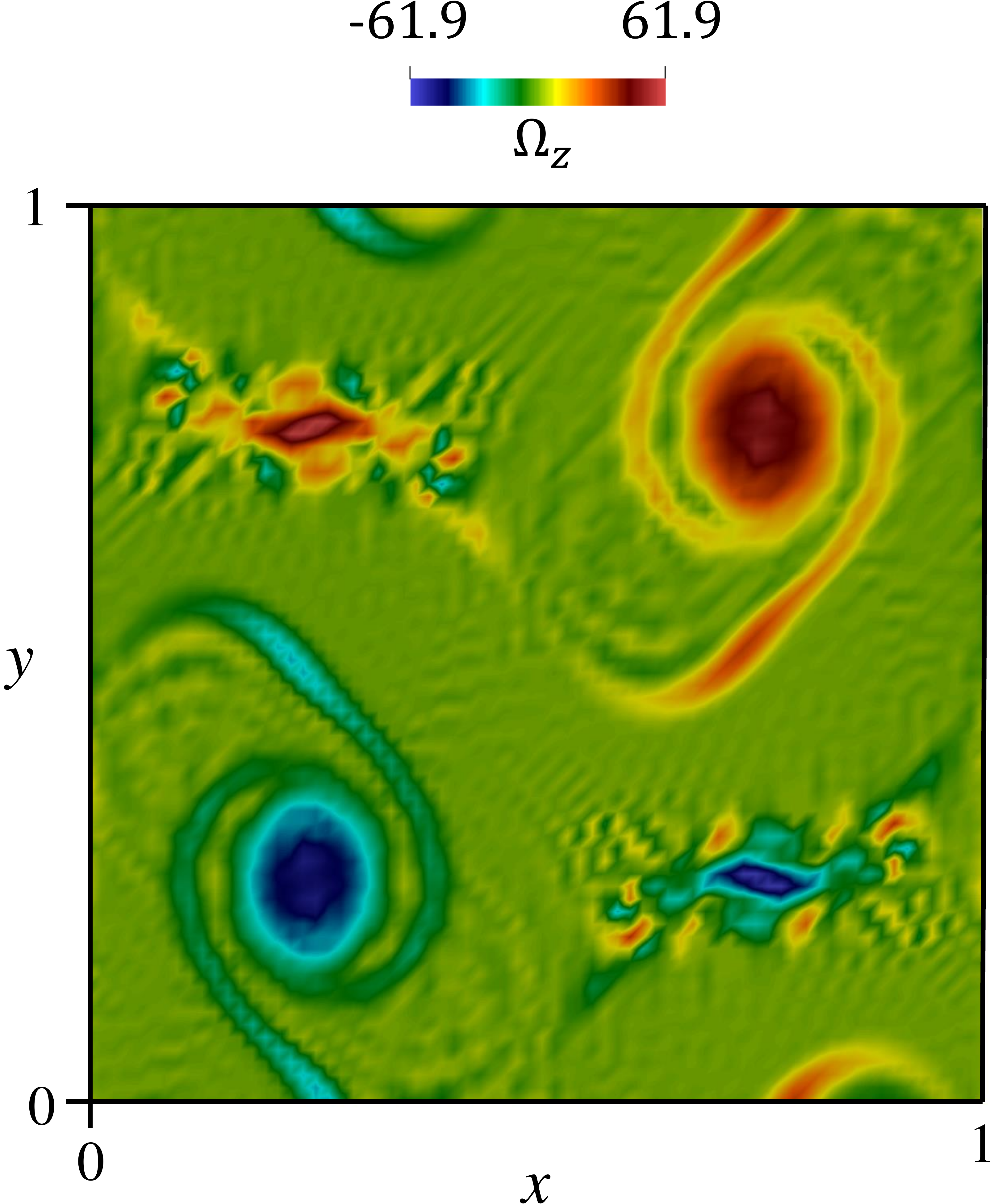} \label{fig:bad_shen}}
\subfigure[NAD-OC6]{\includegraphics[width=0.25\textwidth]{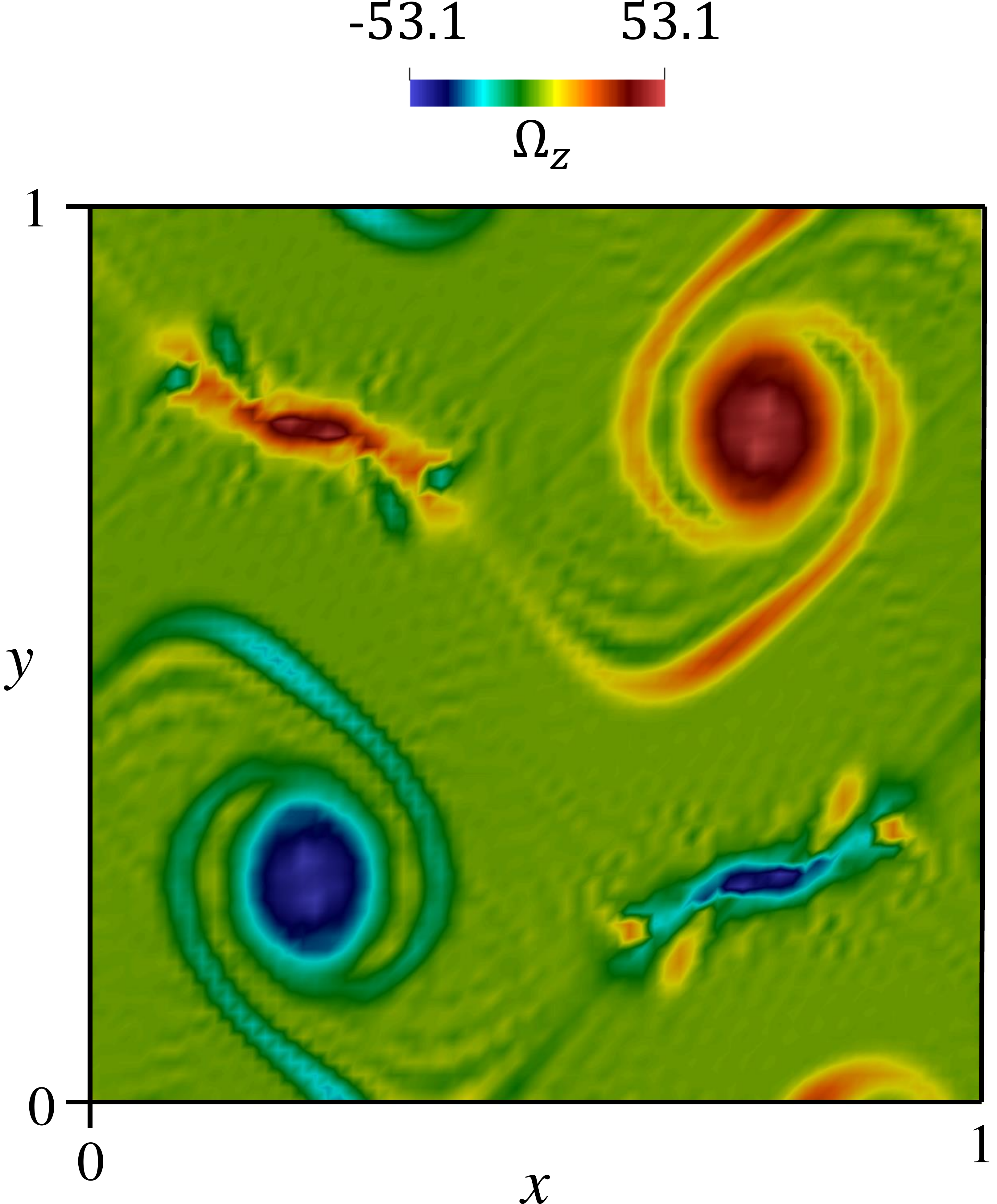}
\label{fig:bad_visbalc6}}
%
\subfigure[$\alpha$-OC6]{\includegraphics[width=0.25\textwidth]{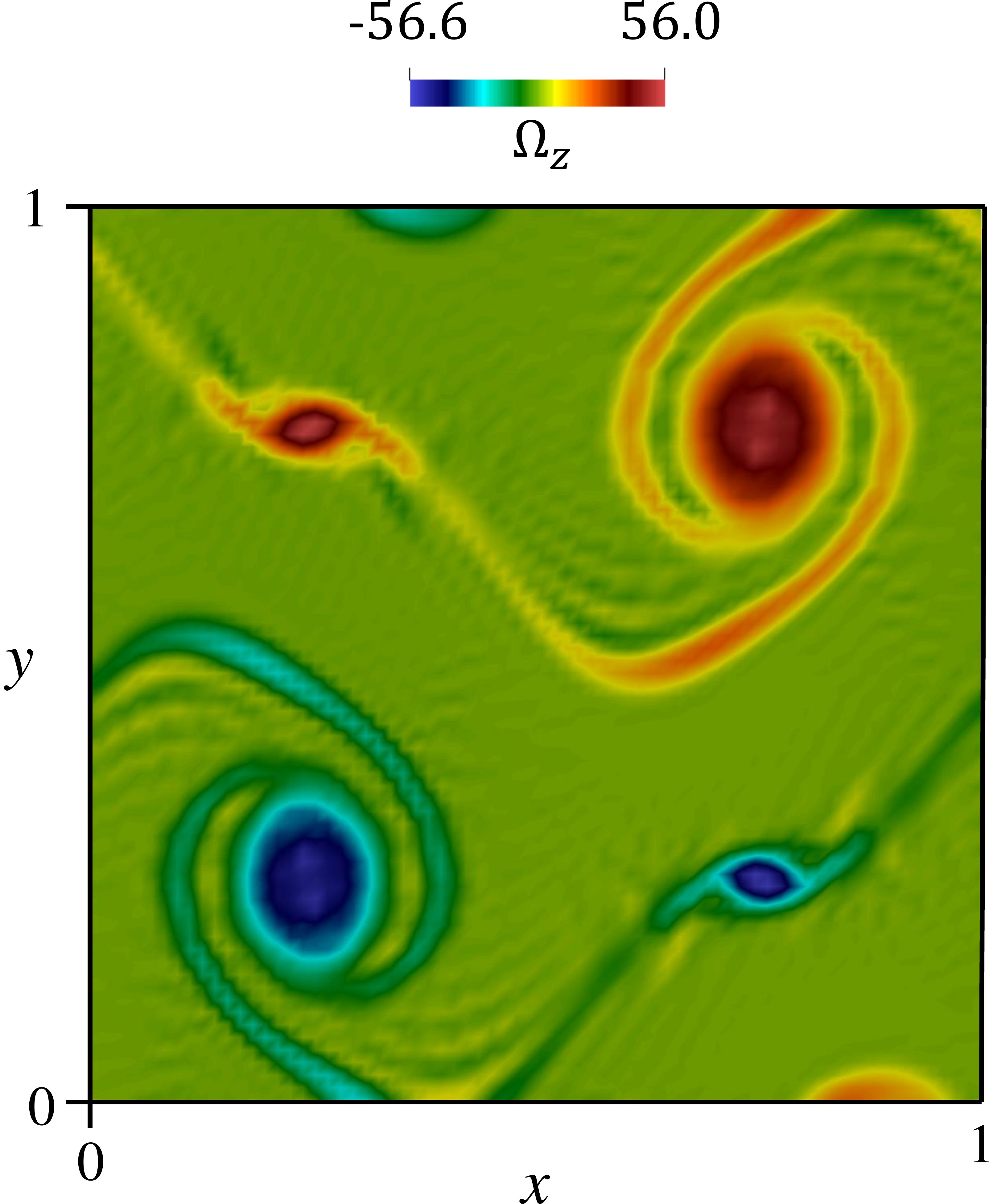} \label{fig:good_aoc6}}
\caption{Vorticity contours of the considered methods on a grid size of $72^2$.}
\label{fig_dpsl_72}
\end{figure}

The reference DNS solution is computed on a grid resolution of $2048^2$ using the $\alpha$-OC6 viscous scheme and is shown in Fig. \ref{DNS_DPSL}. Under-resolved grids, insufficient viscous dissipation, or a combination of the two cause the flow to produce unphysical braid vortices and oscillations. Contours of $z$-vorticity ($\Omega_z$) computed using various viscous flux discretizations are shown at $t=1$ in Figs. \ref{fig_dpsl_72} and \ref{fig_dpsl_144}. The effect of the methods's spectral properties is reflected in their solutions. The solution obtained from a method with superior spectral properties for a large range of wavenumbers is expected to look closer to the DNS solution. \\

\begin{figure}[h!]
\centering
 \includegraphics[width=1.0\textwidth]{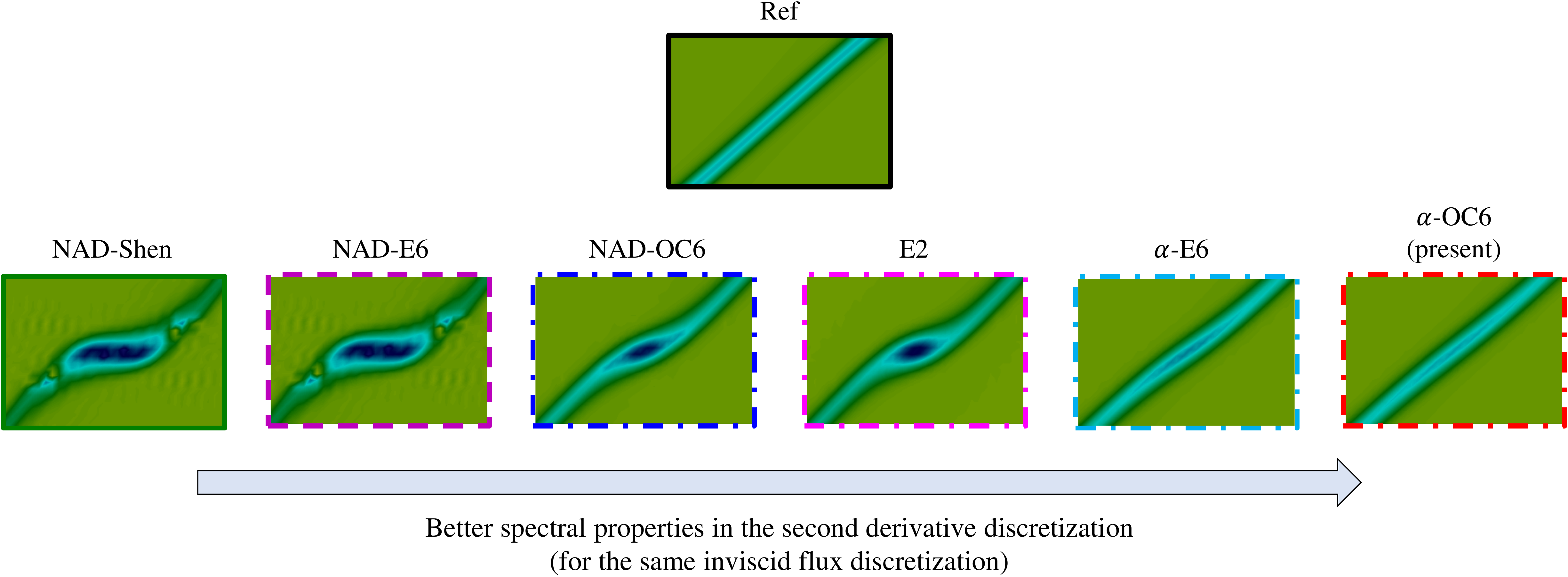}
\caption{Braid vortex of the double periodic shear layer vanishing under the improvement of viscous scheme's spectral properties.}
\label{collage}
\end{figure}

\begin{figure}[h!]

\centering

\subfigure[NAD-Shen]{\includegraphics[width=0.25\textwidth]{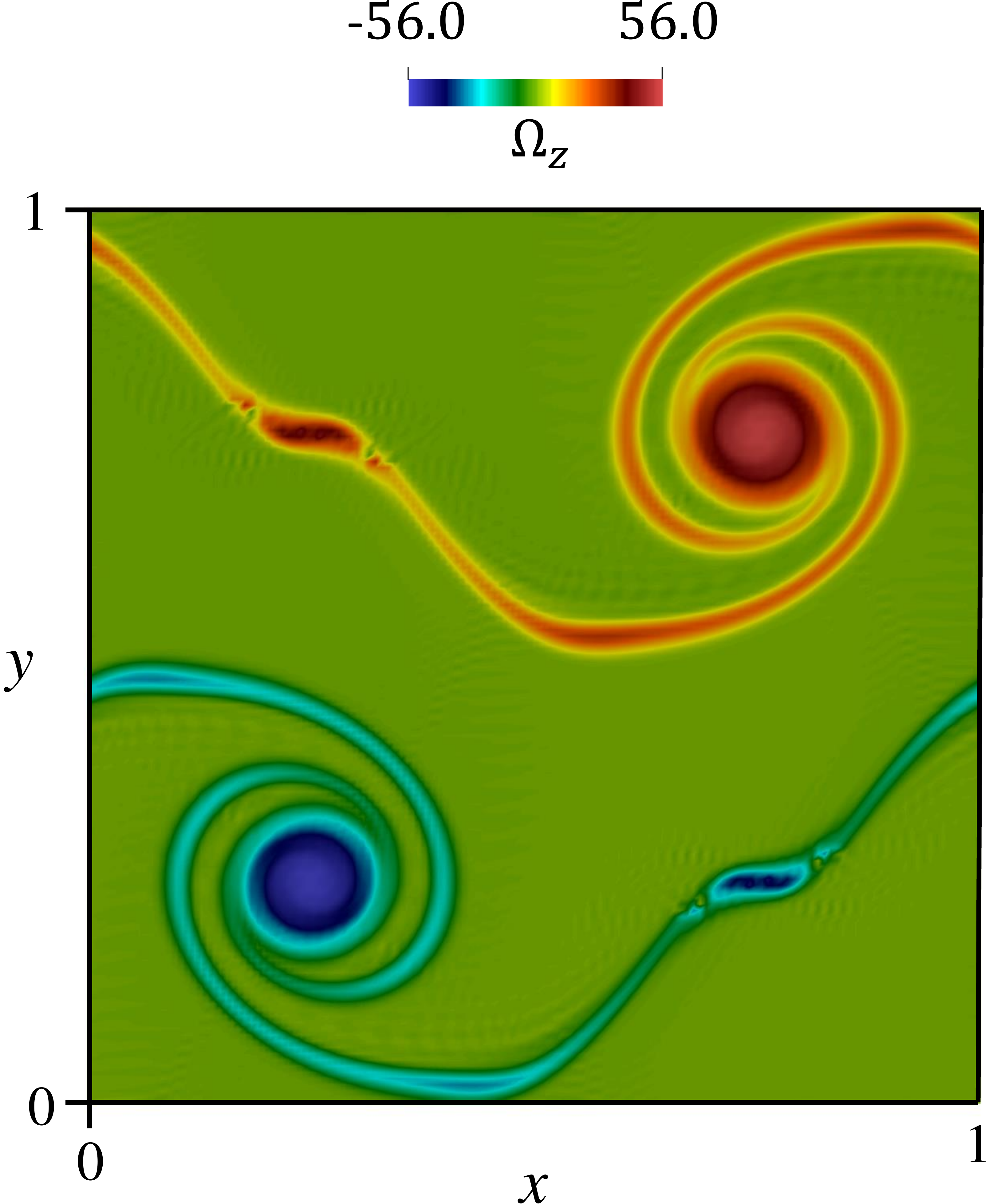} \label{fig:bad_shen_2}}
\subfigure[NAD-E6]{\includegraphics[width=0.25\textwidth]{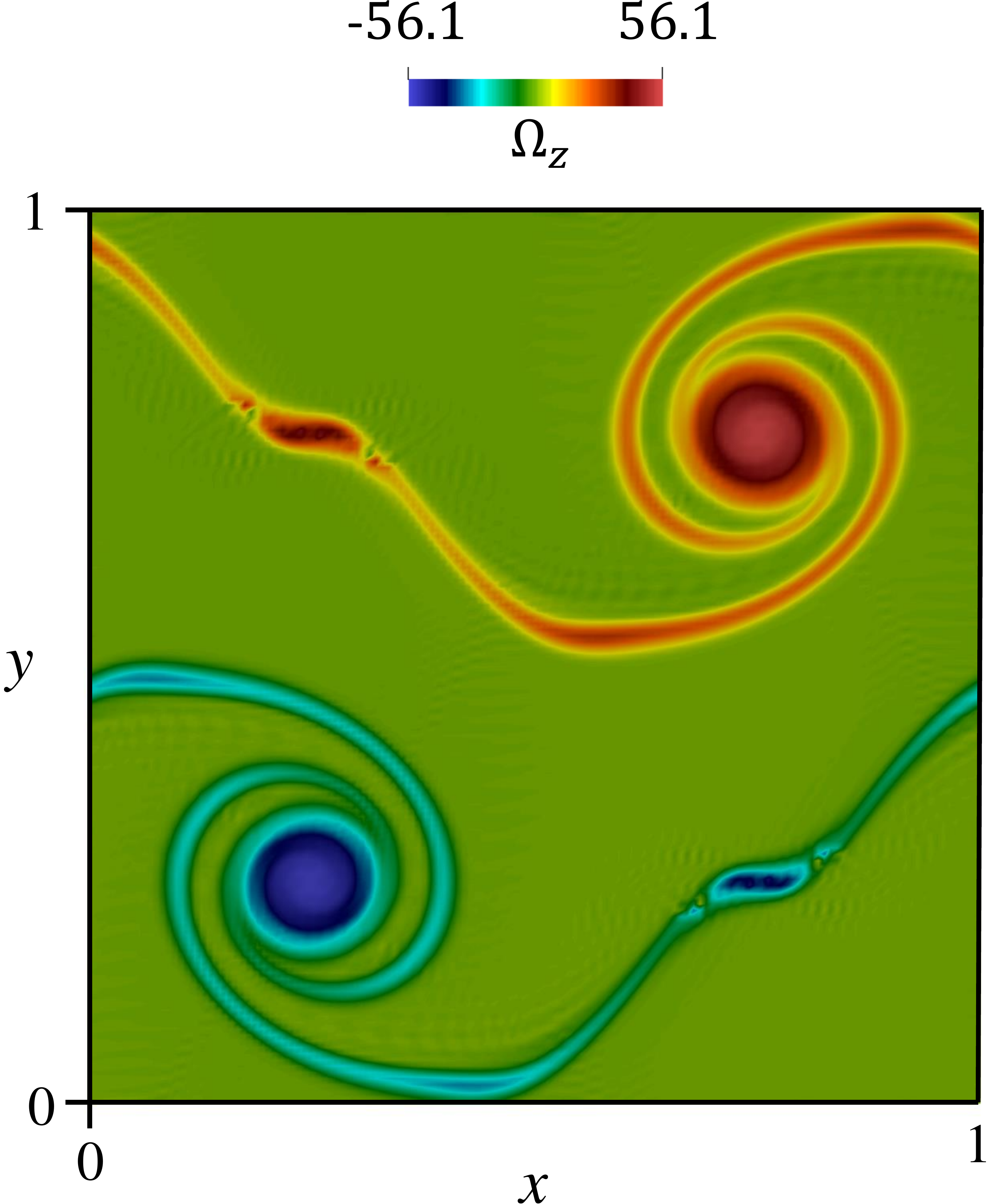}
\label{fig:bad_e62}}
\subfigure[NAD-OC6]{\includegraphics[width=0.25\textwidth]{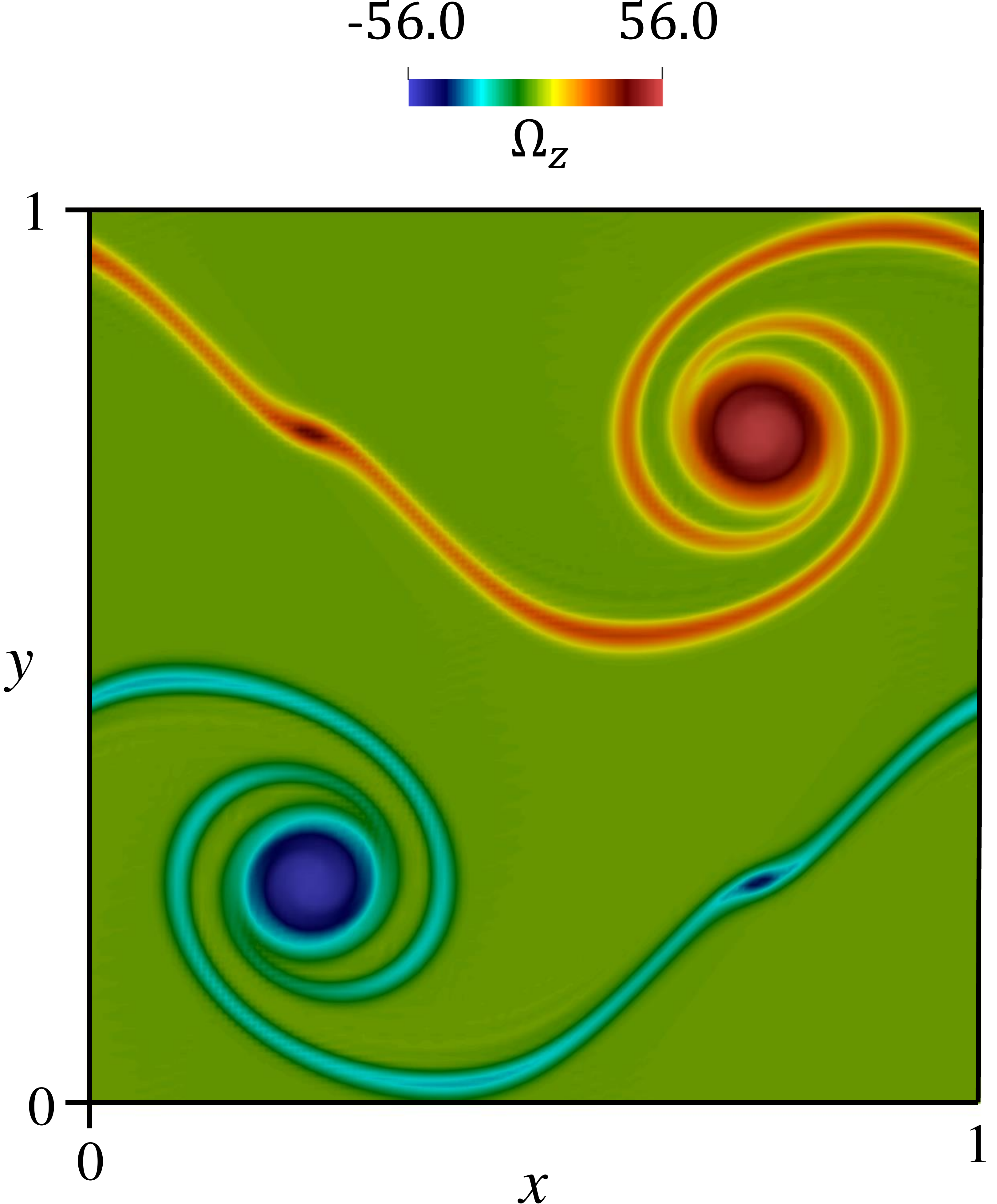}
\label{fig:bad_visbalc62}}

\subfigure[E2]{\includegraphics[width=0.25\textwidth]{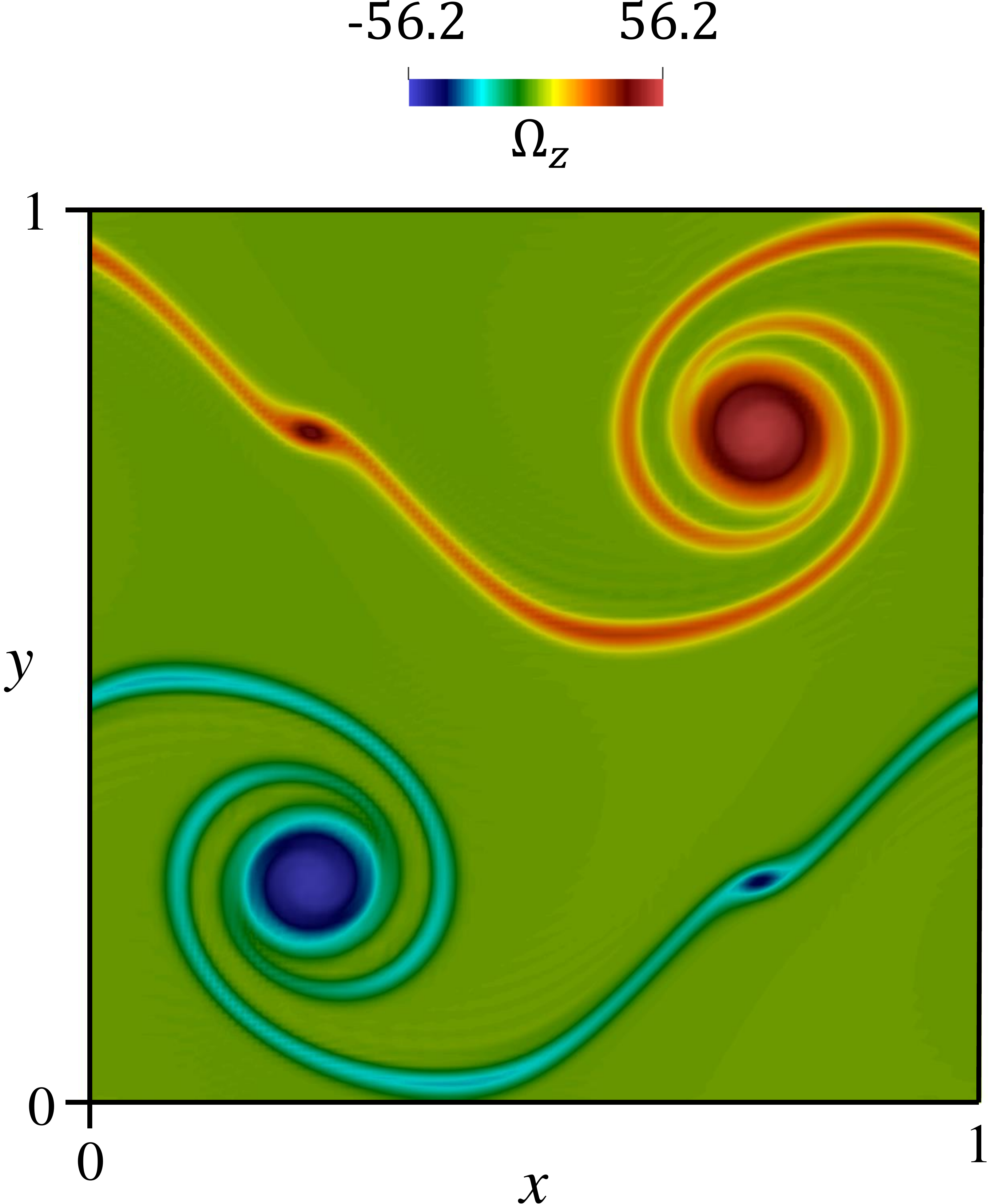}
\label{fig:good_ae2}}
\subfigure[$\alpha$-E6]{\includegraphics[width=0.25\textwidth]{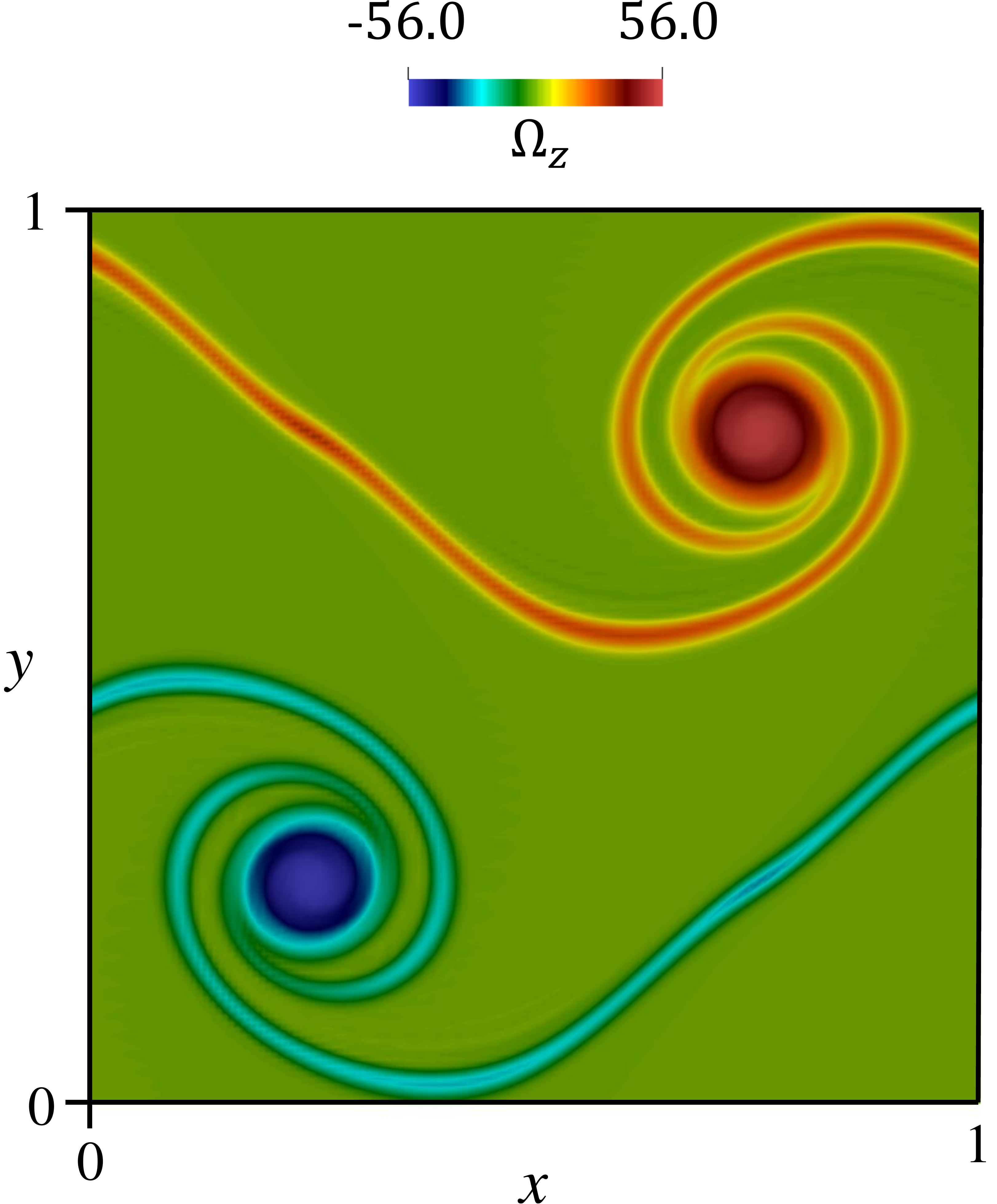}
\label{fig:good_ae62}}
\subfigure[$\alpha$-OC6]{\includegraphics[width=0.25\textwidth]{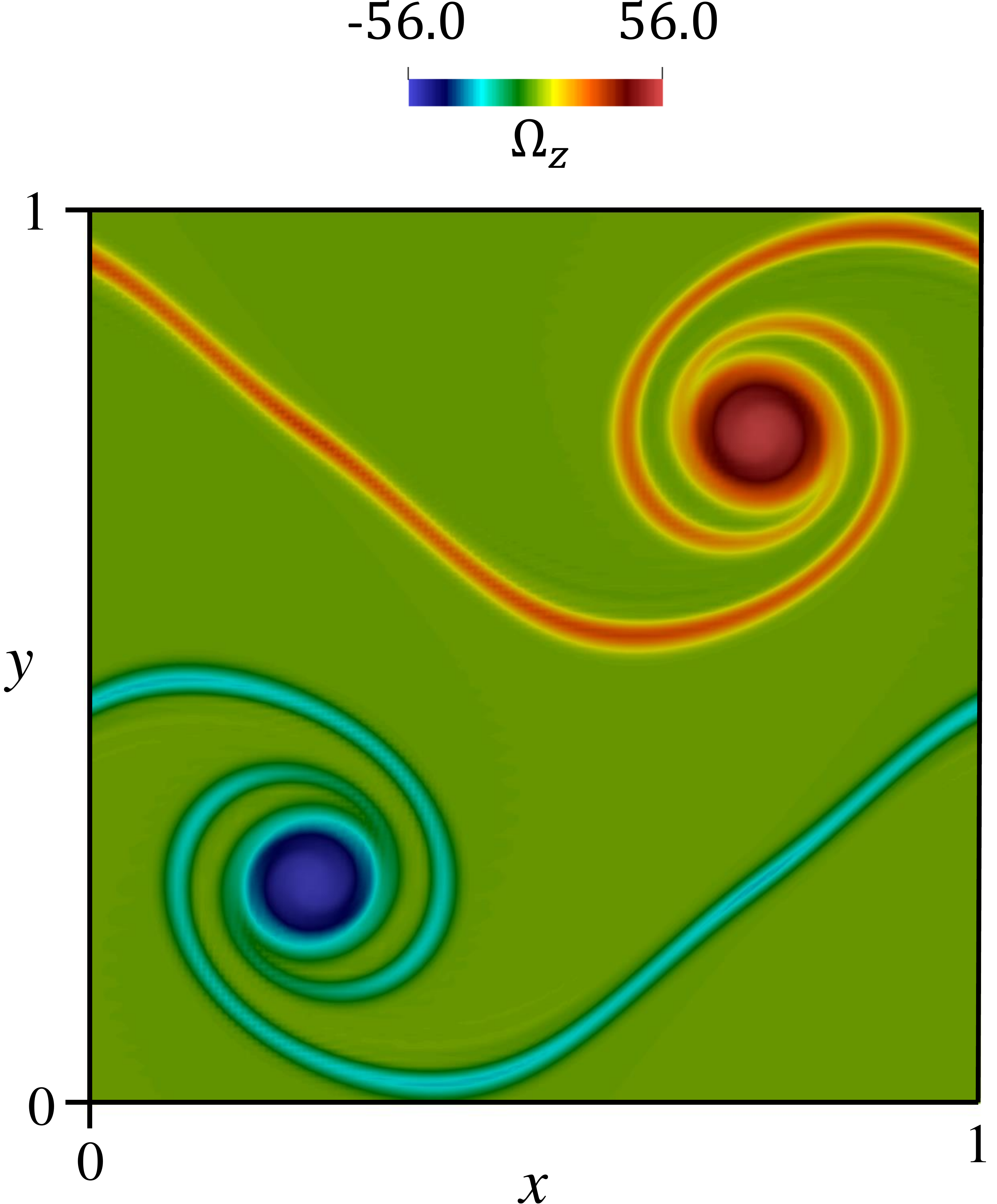} 
\label{fig:good_aoc62}}

\caption{Vorticity contours of the considered methods on a grid size of $144^2$.}
\label{fig_dpsl_144}
\end{figure}

The results shown in Figs. \ref{fig_dpsl_72} and \ref{fig_dpsl_144} are consistent with the spectral plots in Fig. \ref{fig:second_deriv}. These plots show that the oscillations and the braid-vortex features present in them are both minimized with the use of the $\alpha$-OC6 scheme, which has the superior spectral properties among all the candidate schemes considered in this paper. On the other hand, the schemes underestimating the viscous dissipation yielded flow features with extra vortices and non-linear features. It has to be noted that these extra vortices in the flow field do not represent a true solution, as can be seen from the fine grid solution in Fig. \ref{DNS_DPSL}. Oscillations in the low grid resolution of $72^2$ are because the grid employed is incapable of resolving the low wavenumber features where the dissipation occurs. A better visualization summarizing the role of second derivative discretization on the solution can be observed in Fig. \ref{collage}. The schemes that resemble the exact spectral curve are much closer to the DNS result.

\section{\textcolor{black}{Conclusions and Future work}}\label{sec5} 

In this note, we analyzed the spectral properties of various viscous flux discretizations and performed relevant numerical experiments to understand their effect on flow simulations. The NAD methods performed poorly and produced physically inconsistent numerical results compared to the $\alpha$-damping methods, which can be directly attributed to their spectral properties. The magnitude of second derivative values is under-estimated in the NAD viscous methods, which is evident from the dispersion plot. This effect is reflected in the numerical experiments where the NAD viscous methods show features corresponding to this under-estimation. Therefore, they may not be suitable candidates for marginally resolved simulations. \\

In the $\alpha$-damping methods, the second-order central scheme did not provide sufficient damping in the high wavenumber regions and over-predicted the enstrophy values in the viscous TGV simulations. The sixth-order explicit and the newly derived sixth-order compact $\alpha$-damping methods appropriately predicted the enstrophy values for the given grid size, and the results are found to be physically consistent. Contrary to the statement in  Ref. \cite{debonis2013solutions}, viscous flux discretization plays a prominent role in turbulent flow simulations. Viscous flux discretization is not only important for compressible flows with discontinuities, as shown in \cite{chamarthi2022}, but also for shock-free turbulent flows. The $\alpha$-damping approach for viscous fluxes has been extended to multi-species flows in Ref. \cite{chamarthi2022gradient} and is expected to benefit even multiphase and particle-laden flow simulations, which are currently ongoing and will be presented elsewhere.\\

\textcolor{black}{The present study focused on the gradients' computation and spectral properties of the viscous flux discretization. But the viscous fluxes computation can also be affected by the calculation of the diffusion coefficients and time integration. For the test cases considered in the present work, the effect of the diffusion coefficient computation is minimal. Still, it will impact if the diffusion coefficients are large (as in multiphase flows), strongly temperature dependent (as in reacting multi-species flows), and even discontinuous coefficients. Furthermore, the temporal advancement scheme also introduces numerical dissipation, which interacts in an unknown way with the convective and viscous terms. These issues can be considered as future work.}

\section*{Acknowledgements}
The authors gratefully acknowledge the financial support provided by the European Union’s Horizon 2020 Research And Innovation Programme under grant agreement No. 815278. The authors also gratefully acknowledge the support given by the Technion-Israel Institute of Technology in Haifa, Israel. 


\bibliographystyle{elsarticle-num}

\begin{thebibliography}{10}
\expandafter\ifx\csname url\endcsname\relax
  \def\url#1{\texttt{#1}}\fi
\expandafter\ifx\csname urlprefix\endcsname\relax\def\urlprefix{URL }\fi
\expandafter\ifx\csname href\endcsname\relax
  \def\href#1#2{#2} \def\path#1{#1}\fi

\bibitem{debonis2013solutions}
J.~DeBonis, Solutions of the taylor-green vortex problem using high-resolution
  explicit finite difference methods, in: 51st AIAA Aerospace Sciences Meeting
  including the New Horizons Forum and Aerospace Exposition, 2013, p. 382.

\bibitem{chandrashekar2013kinetic}
P.~Chandrashekar, Kinetic energy preserving and entropy stable finite volume
  schemes for compressible euler and navier-stokes equations, Communications in
  Computational Physics 14~(5) (2013) 1252--1286.

\bibitem{kuya2018kinetic}
Y.~Kuya, K.~Totani, S.~Kawai, Kinetic energy and entropy preserving schemes for
  compressible flows by split convective forms, Journal of Computational
  Physics 375 (2018) 823--853.

\bibitem{pirozzoli2010generalized}
S.~Pirozzoli, Generalized conservative approximations of split convective
  derivative operators, Journal of Computational Physics 229~(19) (2010)
  7180--7190.

\bibitem{kennedy2008reduced}
C.~A. Kennedy, A.~Gruber, Reduced aliasing formulations of the convective terms
  within the navier--stokes equations for a compressible fluid, Journal of
  Computational Physics 227~(3) (2008) 1676--1700.

\bibitem{Shu1997}
C.~W. Shu, {Essentially Non-Oscillatory and Weighted Essentially
  Non-Oscillatory Schemes for Hyperbolic Conservation Laws Operated by
  Universities Space Research Association}, ICASE Report~(97-65) (1997) 1--78.

\bibitem{Hu2010}
X.~Y. Hu, Q.~Wang, N.~A. Adams, {An adaptive central-upwind weighted
  essentially non-oscillatory scheme}, Journal of Computational Physics
  229~(23) (2010) 8952--8965.
\newblock \href {https://doi.org/10.1016/j.jcp.2010.08.019}
  {\path{doi:10.1016/j.jcp.2010.08.019}}.

\bibitem{visbal2002use}
M.~R. Visbal, D.~V. Gaitonde, On the use of higher-order finite-difference
  schemes on curvilinear and deforming meshes, Journal of Computational Physics
  181~(1) (2002) 155--185.

\bibitem{nagarajan2003robust}
S.~Nagarajan, S.~K. Lele, J.~H. Ferziger, A robust high-order compact method
  for large eddy simulation, Journal of Computational Physics 191~(2) (2003)
  392--419.

\bibitem{blaisdell1991numerical}
G.~A. Blaisdell, Numerical simulation of compressible homogeneous turbulence,
  Ph.D. thesis, Stanford University (1991).

\bibitem{lele1992compact}
S.~K. Lele, Compact finite difference schemes with spectral-like resolution,
  Journal of Computational Physics 103~(1) (1992) 16--42.

\bibitem{lamballais2011straightforward}
E.~Lamballais, V.~Fortun{\'e}, S.~Laizet, Straightforward high-order numerical
  dissipation via the viscous term for direct and large eddy simulation,
  Journal of Computational Physics 230~(9) (2011) 3270--3275.

\bibitem{dairay2017numerical}
T.~Dairay, E.~Lamballais, S.~Laizet, J.~C. Vassilicos, Numerical dissipation
  vs. subgrid-scale modelling for large eddy simulation, Journal of
  Computational Physics 337 (2017) 252--274.

\bibitem{lamballais2021viscous}
E.~Lamballais, R.~V. Cruz, R.~Perrin, Viscous and hyperviscous filtering for
  direct and large-eddy simulation, Journal of Computational Physics 431 (2021)
  110115.

\bibitem{shen2009high}
Y.~Shen, G.~Zha, X.~Chen, High order conservative differencing for viscous
  terms and the application to vortex-induced vibration flows, Journal of
  Computational Physics 228~(22) (2009) 8283--8300.

\bibitem{shen2010large}
Y.~Shen, G.~Zha, Large eddy simulation using a new set of sixth order schemes
  for compressible viscous terms, Journal of Computational Physics 229~(22)
  (2010) 8296--8312.

\bibitem{chamarthi2022}
A.~S. Chamarthi, S.~Bokor, S.~H. Frankel, On the importance of high-frequency
  damping in high-order conservative finite-difference schemes for viscous
  fluxes, Journal of Computational Physics (2022) 111195.

\bibitem{nishikawa:AIAA2010}
H.~Nishikawa, Beyond interface gradient: A general principle for constructing
  diffusion schemes, in: Proc. of 40th {AIAA} Fluid Dynamics Conference and
  Exhibit, {AIAA} Paper 2010-5093, Chicago, IL, 2010.

\bibitem{chamarthi2022gradient}
A.~S. Chamarthi, Gradient based reconstruction: Inviscid and viscous flux
  discretizations, shock capturing, and its application to single and
  multicomponent flows, arXiv preprint arXiv:2205.01034 (2022).

\bibitem{kuya2021high}
Y.~Kuya, S.~Kawai, High-order accurate kinetic-energy and entropy preserving
  (keep) schemes on curvilinear grids, Journal of Computational Physics (2021)
  110482.

\bibitem{laney1998computational}
C.~B. Laney, Computational gasdynamics, Cambridge university press, 1998.

\bibitem{subramaniam2019high}
A.~Subramaniam, M.~L. Wong, S.~K. Lele, A high-order weighted compact high
  resolution scheme with boundary closures for compressible turbulent flows
  with shocks, Journal of Computational Physics 397 (2019) 108822.

\bibitem{kim1996optimized}
J.~W. Kim, D.~J. Lee, Optimized compact finite difference schemes with maximum
  resolution, AIAA journal 34~(5) (1996) 887--893.

\bibitem{luo2013comparison}
J.~Luo, L.~Xuan, K.~Xu, Comparison of fifth-order weno scheme and finite volume
  weno-gas-kinetic scheme for inviscid and viscous flow simulation,
  Communications in Computational Physics 14~(3) (2013) 599--620.

\bibitem{Jiang1995}
G.-S. Jiang, C.-W. Shu, {Efficient Implementation of Weighted ENO Schemes},
  Journal of Computational Physics 126~(126) (1995) 202--228.

\bibitem{brachet1983small}
M.~E. Brachet, D.~I. Meiron, S.~A. Orszag, B.~Nickel, R.~H. Morf, U.~Frisch,
  Small-scale structure of the taylor--green vortex, Journal of Fluid Mechanics
  130 (1983) 411--452.

\bibitem{abe2018stable}
Y.~Abe, I.~Morinaka, T.~Haga, T.~Nonomura, H.~Shibata, K.~Miyaji, Stable,
  non-dissipative, and conservative flux-reconstruction schemes in split forms,
  Journal of Computational Physics 353 (2018) 193--227.

\bibitem{bassi1997high}
F.~Bassi, S.~Rebay, A high-order accurate discontinuous finite element method
  for the numerical solution of the compressible navier--stokes equations,
  Journal of computational physics 131~(2) (1997) 267--279.

\end{thebibliography}

\end{document}